\documentstyle[aps,amsmath,amssymb,epsfig]{revtex}

\begin{document}

\bibliographystyle{plain}

\title{The two-boundary sine-Gordon model}

\author{J.-S. Caux$^1$, H. Saleur$^{2,3,4}$ and F. Siano$^5$}

\address{$^1$Instituut voor Theoretische Fysica, Universiteit van 
Amsterdam, 1018 XE Amsterdam, The Netherlands\\
$^2$ Department of Physics and Astronomy, University of Southern
California, Los Angeles, CA 90089 \\
$^3$ Service de Physique Th\'eorique, CEA Saclay, Gif Sur Yvette 91191, France\\
$^4$ Physikalisches Institut, Albert-Ludwigs-Universit\"at, D-79104, Freiburg, Germany\\
$^5$ Institut f\"ur Theoretische Physik, Heinrich-Heine Universit\"at, 
D-40225 D\"usseldorf, Germany} 

\maketitle

\begin{abstract}
We study in this paper the ground state energy of a free bosonic theory on a finite 
interval of length $R$ with either a pair of sine-Gordon type or a pair of Kondo type 
interactions at each boundary. This problem has potential applications in condensed 
matter (current through superconductor-Luttinger liquid-superconductor junctions) 
as well as in open string theory (tachyon condensation). While the application of 
Bethe ansatz techniques to this problem is in principle well known, considerable 
technical difficulties are encountered. These difficulties arise mainly from the way 
the bare couplings are encoded in the reflection matrices, and require complex 
analytic continuations, which  we carry out in detail in a few cases. 
\end{abstract}


\section{Introduction}

This paper is concerned with the study of $1+1$ quantum field
theories defined on a finite interval of length $R$. When the bulk is
massless and conformal boundary conditions are chosen, such theories 
can be entirely understood using the formalism of boundary conformal 
field theory \cite{CardyBOOK}. The case where one deviates from
this situation either by adding an interaction in the bulk or at the 
boundaries (or both) is more difficult, and quite interesting as it
involves crossovers depending on the bulk and boundary mass scales as
well as the finite size $R$. 

Since the ``conformal revolution'', 
this topic was probably first considered in print in \cite{LeClairNPB453}
(although it has been studied in unpublished work of A.
Zamolodchikov) where the basic thermodynamic Bethe ansatz and Destri 
de Vega (DDV) approaches were delineated in the case of simple
integrable theories, mostly massive in the bulk, with simple boundary
conditions. Since then, the topic has attracted a fair amount of
attention (see for instance \cite{Itoyama}, \cite{DoreyNPB525},\cite{Feverati}, 
\cite{Takacs},\cite{Giuseppe}).

In the present  paper, we discuss in details the limit complementary to 
\cite{LeClairNPB453} where the bulk is a massless theory - mostly 
a free boson - and the perturbations sit on the boundary. We shall
have mostly two theories in mind. The first is of the boundary
sine-Gordon model type \footnote{In \cite{LeClairNPB453} the
results referred to as boundary sine-Gordon model concern mostly a
bulk sine-Gordon theory with Dirichlet type boundary conditions in
$0,R$. The terminology has somewhat changed since then, and
boundary sine-Gordon theory nowadays refers rather to a free boson 
with a cosine perturbation at the boundary.}
\begin{eqnarray}
H = \frac{1}{2} \int_0^R dx [(\partial_t \phi)^2 + (\partial_x
\phi)^2]+\tilde{\Delta}_l \cos \frac{\beta}{2} \phi(0) +
\tilde{\Delta}_r \cos \left(\frac{\beta}{2} \phi(R) - 
\chi\right)\label{basic1},
\end{eqnarray}
and the second of Kondo type
\begin{eqnarray}
H = \frac{1}{2} \int_0^R dx [(\partial_t \phi)^2 + (\partial_x
\phi)^2] + \tilde{\Delta}_l \left[e^{i\frac{\beta}{2} \phi(0)}S_{l}^{-}+
e^{-i\frac{\beta}{2} \phi(0)}S_{l}^{+}\right]
+ \tilde{\Delta}_r \left[e^{i\frac{\beta}{2} \phi(R)}S_{r}^{-}+
e^{-i\frac{\beta}{2} \phi(R)}S_{r}^{+}\right]\label{basic2}
\end{eqnarray}
In this last equation, $S_{l,r}^{+,-}$ refer to $U_{q}sl(2)$ spins in
representations of spin $j_{l},j_{r}$ and $q=e^{-i\pi\beta^{2}}$. The
case with $j_{l}=j_{r}=1/2$ could be called a ``double Kondo problem''.
In that case, as usual,  the $\beta$ dependence of the interaction
can be traded for a coupling anisotropy in spin space - notice that
we restrict to the case where the $l$ and $r$ sides see the same
anisotropy, a  condition necessary to preserve integrability. 

This double Kondo theory musn't be confused with the two impurtity
Kondo model. Although there also one would expect results to depend
on the couplings for each impurity and the distance between the
impurities, the detailed set-up is entirely different \cite{Affleck}.

There are plenty of physical motivations to tackle these problems.
On the condensed matter side, they can be thought of as 
generalizations of quantum impurity problems, where the bulk
of the theory gets modified by the addition of a single local defect,
e.g. a boundary or a quantum degree of freedom.  Now, instead of only
one impurity, we have two different ones coupled by a finite-size
quantum field theory.  One could label such a situation as a ``quantum
bridge problem''. An experimental setup for such problems  could
involve 
an interacting device - maybe a
carbon nanotube or quantum wire - connected to two leads on either end.
If the excitations of the device suffer from a gap when trying to
infiltrate the leads, then in a first approximation all the action
takes place at the contacts between the device and the leads.  The
excitations, upon hitting the contacts, will be reflected and mixed by
the ``boundary'' according to some rules dictated by the specifics of
the couplings, in other words the excitations will face some 
nontrivial, energy dependent boundary conditions, and the situation
could be described by hamiltonians such as
(\ref{basic1},\ref{basic2}) in the low energy, universal limit. 

Possibly the simplest example of such a situation occurs when one
considers two different (s-wave) superconductors connected by a bridge
made of a quantum wire, e.g. a carbon nanotube, a construction called
a Josephson junction.  Since both superconductors inherently possess a gap for electronic
excitations within their bulk, an effective theory can be obtained by
considering only the wire itself with two ``impurities'' on either
side replacing the superconductors.  Solving this effective model -
which is essentially described by (\ref{basic1}) 
leads to the determination of the 
Josephson current \cite{CauxPRL88}.

We do not know of any experimental condensed matter  situation where the two boundary  Kondo
model would be relevant (although this does not seem impossible to imagine). 
The model has however appeared in string theory in the study of tachyon instabilities \cite{Bardakci} where the ends of the open string are coupled to a background, and the boundary spin acts in the space of Chan-Paton factors (and the free boson corresponds to the $X^{25}$ coordinate). The double boundary sine-Gordon model could presumably also be interpreted in the context of tachyon condensation following \cite{Harvey}.

Notice that in the hamiltonian (\ref{basic2}) we have not introduced 
the phase difference $\chi$. This is because such a phase can always 
be absorbed into a (gauge) redefinition of the spin operators
$S^{\pm}$. Alternatively, observe that a perturbative computation of 
the ground state energy of this model will only involve
configurations of charges (the charges in the boundary vertex
operators) which are independently neutral on the left and right
sides, which leads to the possibility of shifting $\phi$ by a
constant independently on each boundary (notice also that this ground
state energy will then expand in powers of
$\Delta_{l}^{2},\Delta_{r}^{2}$. The situation for
(\ref{basic1}) is very different, as all charge configurations
which are only neutral overall (ie by combining left and right sectors) do
contribute. The dependence of the ground state energy on $\chi$ is
actually one of the main concerns of this paper. 

Let us now give some generalities about our approach. We
consider a strip of width $R$ and length $L \rightarrow
\infty$, with Euclidean coordinates $x \in [0,R]$ and $y \in {\Bbb R}$.  
We will mostly discuss the double boundary sine-Gordon model in this paper (\ref{basic1}), which belongs to a more 
general class of bulk and boundary perturbed conformal field
theories, whose Euclidian action can be schematically written 
\begin{eqnarray}
S = S_{CFT + CBC} + \int_{0}^R dx \int dy ~\Phi(x,y) + \int dy \left
( \Phi_{B_l} (y) + \Phi_{B_r} (y) \right).
\label{S2}
\end{eqnarray}
Here,we have  as a starting point a  CFT with action
$S_{CFT}$, perturbed by some relevant operator $\Phi(x,y)$ contained
within the operator content of that particular CFT.  In the presence
of boundaries, one needs to specify conformal boundary conditions
(CBCs) at $x = 0,R$, and include the nontrivial boundary effects as a
perturbation on the CBCs by some relevant boundary operators (in which we have included 
coupling constants $g_l,g_r$)
$\Phi_{B_{l,r}} (y) \equiv g_{l,r} \Phi (x,y)|_{x=0,R}$ at the left and right
boundaries.  

The theory (\ref{basic1})  turns out to be
integrable. This can be shown by a simple generalization of
the argument of \cite{GhoshalIJMPA9}, which we discuss here for
completeness. 
 The Lagrangian or perturbed CFT
approaches both provide a symmetric stress-energy tensor $T_{\mu \nu}$
with components $T_{zz} \equiv T, T_{z \bar{z}} \equiv \Theta$ 
satisfying the continuity equations
\begin{eqnarray}
\partial_{\bar{z}} T = \partial_z \Theta, \hspace{3cm} 
\partial_z \bar{T} = \partial_{\bar{z}} \bar{\Theta}.
\label{continuity}
\end{eqnarray}
In the presence of boundaries, as was shown by Cardy \cite{CardyBOOK},
the choice of conformal boundary conditions ensures that no momentum
can flow through the boundaries, in other words that the off-diagonal
components of the stress-energy tensor vanish at the boundaries,
$T_{xy} |_{x=0,R} =0$.
In the presence of boundary perturbations, however, this equation gets
modified to
\begin{eqnarray}
T_{xy} |_{x=0,R} = -i (T-\bar{T})|_{x=0,R} = \mp \frac{d}{dy}
\vartheta_{l,r} 
\label{boundaryT}
\end{eqnarray}
where $\vartheta$ is some local boundary field, which can be related
to the boundary fields $\Phi_{B_{l,r}}$.

\begin{figure}
\centerline{\epsfig{figure=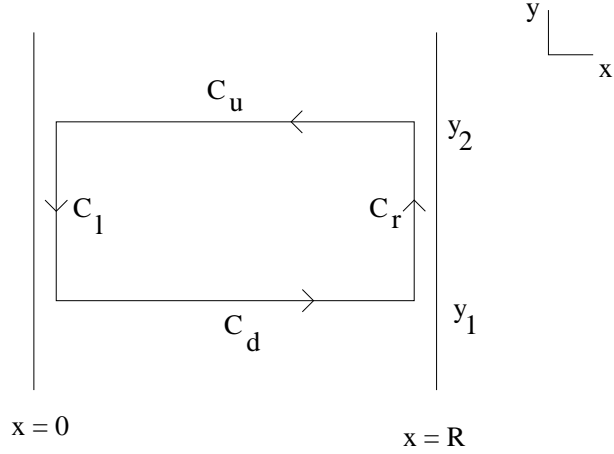,width=8cm}}
\vspace{0.3cm}
\caption{Contour of integration for the construction of the integrals
of motion.}
\label{FirstContour}
\end{figure}

Consider now the integrals over a closed contour $C$ 
\begin{eqnarray}
P_1(C) = \int_C (T dz + \Theta d\bar{z}), \hspace{3cm} \bar{P} (C) =
\int_C (\bar{T} d\bar{z} + \bar{\Theta} dz).
\end{eqnarray}
By the continuity equations (\ref{continuity}), these do not change
under deformations of the contour $C$:  $P_1(C) = \bar{P}_1 (C) =0$.
Let us now take $C$ to be the contour illustrated in figure
\ref{FirstContour}. 
We then can write
\begin{eqnarray}
0 = P_1 (C) + \bar{P}_1 (C) = \sum_{a = u,d,l,r} P_1 (C_a) + \bar{P}_1
(C_a). 
\end{eqnarray}
The contours $C_l$ and $C_r$, by equation (\ref{boundaryT}), can be
seen to give the contributions
\begin{eqnarray}
P_1 (C_{l,r}) + \bar{P}_1 (C_{l,r}) = \vartheta_{l,r} (y_1) -
\vartheta_{l,r} (y_2).
\end{eqnarray}
If we take $y$ to be the time direction of our quantization scheme,
this then shows that the quantity
\begin{eqnarray}
H_{l,r} (y) = \int_0^R dx (T + \bar{T} + 2 \Theta) + \vartheta_l (y) +
\vartheta_r (y)
\label{FirstHamiltonian}
\end{eqnarray}
is $y$-independent, in other words is an integral of motion.

If the theory is integrable, these equations are but the first
elements of an infinite series.  Then, there are an infinite number of
local fields $T_{s+1}, \Theta_{s-1}$ obeying
\begin{eqnarray}
\partial_{\bar{z}} T_{s+1} = \partial_z \Theta_{s-1}, \hspace{3cm}
\partial_{\bar{z}} \bar{T}_{s+1} = \partial_z \bar{\Theta}_{s-1}
\end{eqnarray}
in the bulk.  Provided they obey the boundary conditions
\begin{eqnarray}
(T_{s+1} + \Theta_{s-1} - \bar{T}_{s+1} -\bar{\Theta}_{s-1})|_{x=0,R}
= \mp i\frac{d}{dy} \vartheta_{s; l,r} (y)
\label{integrableBCs}
\end{eqnarray}
for some infinite subset $\{ s_B \} \in \{ s \}$, the theory will
remain integrable despite the boundaries, with integrals of motion
\begin{eqnarray}
H_{l,r}^{(s)} (y) = \int_0^R dx \left[ T_{s+1} + \Theta_{s-1} +
\bar{T}_{s+1} + \bar{\Theta}_{s-1} \right] + \vartheta_{s;l} (y) +
\vartheta_{s; r} (y).
\end{eqnarray}
Equations (\ref{integrableBCs}) define the possible integrable
boundary conditions.  They formed the fundamental set of conditions
used in the analysis of \cite{GhoshalIJMPA9}.

Integrability can be used in two different - and equivalent through
crossing - ways to calculate the ground state energy. To see this,
let us switch to a hamiltonian formalism; imaginary time can run
along $x$ or $y$.  For time flowing along $y$, it is
natural to identify the first element of the infinite sequence of
conserved quantities, equation (\ref{FirstHamiltonian}), with the
Hamiltonian.  The boundaries are thus true boundaries in space, and
the bulk excitations satisfy appropriate boundary conditions.  In this
language, baptized the $L$-channel, the partition function is obtained
by tracing over all such states, with $L \rightarrow \infty$ as the
inverse temperature:
\begin{eqnarray}
Z = Tr ~e^{-L H_{l,r}}.
\end{eqnarray}
On the other hand, in the   $R$-channel, we take time to flow
along $x$, from an initial state to a final one (i.e. from $x=R$ to
$x=0$).  The Hamiltonian in the bulk is thus the usual one (without
boundary contributions), and the associated space of states is the
same as in the absence of boundaries.  The partition function now
becomes a simple matrix element, obtained by sandwiching the
time-evolution operator $H$ along time $R$ between the initial and
final states (the ``boundary'' states) $|B_{l,r} \rangle$, i.e.
\begin{eqnarray}
Z = \langle B_l | e^{-R H} | B_r \rangle,
\label{RchannelZ}
\end{eqnarray}

This latter point of view leads to the TBA in the crossed channel,
which we will use extensively in what follows. To handle
(\ref{RchannelZ}) one uses the description of the integrable theory as
a massive (if there is also a bulk perturbation) or massless (if the 
perturbation affects only the boundary) scattering theory. Such a
theory is described by the spectrum of particles  together with the bulk and
boundary scattering matrices, and the constraint of factorized
scattering. 

Consider for instance the massive case, and parametrize energy and
momentum of particles through a rapidty $\theta$. The explicit form 
of the boundary states is \cite{GhoshalIJMPA9} 
\begin{eqnarray}
|B\rangle = \exp \left( \int_0^{\infty} d\theta K^{ab} (\theta)
 A^{\dagger}_a (-\theta) A^{\dagger}_b (\theta) \right) | 0 \rangle
\end{eqnarray}
where the amplitudes appearing in the exponential are given by the
analytic continuation of the reflection matrices:
\begin{eqnarray}
K^{ab} (\theta) = R^b_{\bar{a}} \left( \frac{i\pi}{2} - \theta
\right). 
\end{eqnarray}

Using this basic ingredient, a straightforward  method to compute the
ground state energy was proposed in \cite{LeClairNPB453}, in
the case where bulk and boundary scattering are diagonal. The non
diagonal
cases require more work, and discussing them is part of our purpose
here. It so turned out however that even in the diagonal case,
considerable surprises are encountered - as was first discovered in  
\cite{DoreyNPB525}  - due in part to the way the bare couplings are
encoded in the boundary scattering description. Calculating the correct 
ground state energy requires analytical continuation of the results of 
\cite{LeClairNPB453}, a process which is quite difficult.  

The case of double Kondo model (\ref{basic2}) is a bit different, as there are 
additional  boundary degrees of freedom. The proof of integrability would 
proceed along similar lines, with appropriate modifications as discussed 
in \cite{Baseilhac}. The boundary state formalism can also be generalized 
to this case.  We will, however, take a different approach based on the 
Destri-de Vega formalism.
 
Most of this paper is devoted to the double boundary sine-Gordon model 
(sections 2 to 6). We will also present results for the double Kondo model 
in section 7. We shall first
discuss analytic continuation  by considering the  case 
of the Ising model in section 2. In sections 3 and 4, we extend the analysis 
of \cite{LeClairNPB453} to the case where the boundary scattering is non 
diagonal while the bulk scattering is still diagonal. The analytical 
continuation is carried out as a simple generalization of section 2 for the 
case $\beta^2=4\pi$ (many of the results of this section have been previously
published in \cite{CauxPRL88}, but we include a more thorough explanation here,
which helps to understand the generalizations we propose). 
The other ``reflectionless cases'' require considerably 
more work; only the case $\beta^2=\frac{8\pi}{3}$ is fully discussed, in section 5. 
The double 
Kondo model is tackled in section 6; there, the bulk and boundary interactions 
are non diagonal, and we use a different method, based on the
often called Destri-de Vega approach 
\cite{DestriDeVega} (although the method has been used before in slightly 
different settings \cite{Klumper}).

\section{A scalar theory:  the Ising model}

\begin{figure}
\centerline{\epsfig{figure=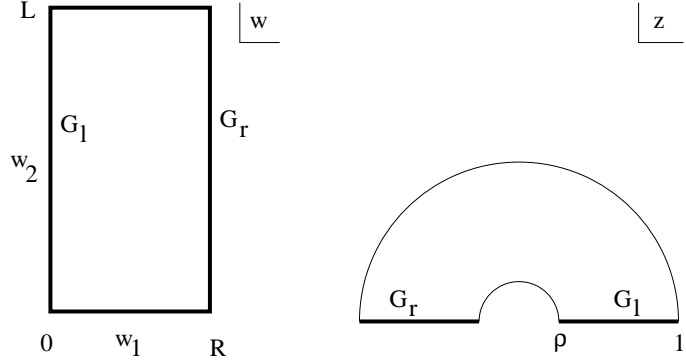,width=9cm}}
\vspace{0.3cm}
\caption{The conformal map between the original strip of $u$ and the complex
plane of $z$.}
\label{Mapping}
\end{figure}

Let us now turn to a specific example, and consider the Ising model.
As is well-known, the bulk action of this theory takes the form of a
free action for Majorana fermions $\psi, \bar{\psi}$.  
At the left and right boundaries
$x=0,R$, there are two (in general different) values of the boundary
magnetic field $h_{l,r}$ coupled to the local boundary spin operators
\begin{eqnarray}
\Phi_{l,r} = \frac{1}{2} (\psi + \bar{\psi})_{x=0,R} ~a_{l,r} (y).
\end{eqnarray}
where $a_{l,r}$ are boundary fermions, and we have chosen a convenient
normalization.  The total action is
\begin{eqnarray}
S = \int_0^R dx \int dy \left[ \psi \partial_{\bar{z}} \psi -
\bar{\psi} \partial_z \bar{\psi} \right] + \int dy
({\cal L}_l + {\cal L}_r), \nonumber \\
{\cal L}_{l,r} = \frac{1}{2} \psi \bar{\psi}|_{x=0,R} + a_{l,r}
\dot{a}_{l,r} + h_{l,r} \Phi_{l,r}.
\end{eqnarray}
To proceed, let us do some elementary perturbative calculations, and adopt 
for a moment a more general setting. 
Consider thus a theory that  is 
conformal in the bulk, and is subject to  boundary perturbations at the left and right 
edges with coupling constants
$g_{l,r}$.  We can write the partition function of our system as 
\begin{eqnarray}
Z = \langle 0 | {\cal T}_y \exp \int_0^L dy \left[ g_l \Phi (0,y) +
g_r \Phi (R,y) \right] | 0 \rangle
\end{eqnarray}
where $| 0 \rangle$ is the ground-state of the unperturbed system,
${\cal T}_y$ is the $y$-ordering operator, and all the correlation
functions are evaluated in the unperturbed CFT with the chosen CBCs.
We introduce a conformal mapping by defining the new coordinates 
\begin{eqnarray}
w = w_i + i w_2 = e^{i \pi z/R}
\end{eqnarray}
in such a way that the original strip maps to the half-disc, as
illustrated in figure \ref{Mapping}.
The boundary is the horizontal axis, the perturbation for $x=0$ now sitting 
on the boundary $w_1>0$, and the one for $x=R$  sitting on the one for
$w_1<0$.
Introducing $d$ the dimension of the operators $\Phi$ (such that
the two point function on the boundary goes as $1/|w|^{2d}$) and
$\rho=e^{-\pi L/R}$, we find 
\begin{equation}
Z_0=<0|{\cal T}\exp\int_\rho^1 \frac{dw}{w^{1-d}} \left(\frac{R}{\pi}\right)^{1-d}\left[g_l \Phi(w)+g_r
\Phi(-w)\right]|0>.
\end{equation}

The strategy is to evaluate $\ln Z_0$ for large $L$, and extract the
leading term proportional to $L$, since in that limit one has 
$\frac{\ln Z_0}{L}\approx -E_0$. Elementary calculations give then
\begin{equation}
E_0=-\frac{\pi c}{24 R} -\frac{\pi}{R}\left(\frac{R}{\pi}\right)^{2(1-d)}\left[(g_l^2+g_r^2) 
\frac{\Gamma(1-2d)\Gamma(d)}{\Gamma(1-d)}+2 \frac{g_l g_r}{d}
F(2d,d;d+1,-1)+...\right].
\label{firstgene} 
\end{equation}
This formula will be useful in what follows.

Let us get back now to the particular case of the Ising model
and  consider the
simplest case where one of the boundary fields vanishes, the other
being equal to $h$. We then obtain  
\begin{equation}
E_0(h)=-\frac{\pi}{48 R}-\frac{\pi}{R} C
\end{equation}
where
\begin{equation}
C=\lim_{\rho\rightarrow 0}{1\over\ln 1/\rho}\sum_{N=1}^\infty
(\sqrt{2}h)^{2N}\int_{\rho\leq |x_1|\leq ... \leq |x_{2N}|\leq 1} 
\left<\Phi(x_1)\ldots\Phi(x_{2N}\right>_c {dx_1\over x_1^{1/2}}\ldots
{dx_{2N}\over x_{2N}^{1/2}} 
\end{equation}
in which $<...>_c$ means that the correlators are evaluated in the
unperturbed CFT.  
Evaluation to first two non trivial orders gives
\begin{equation}
E_0(h)=-{\pi\over 48 R}-\frac{h^2}{4\pi}\sum_{0}^\infty {1\over n+1/2}+ {h^4
R\over{8\pi^2}} \sum_{0}^\infty {1\over(n+1/2)(p+1/2)(n+p+1)} 
\end{equation}
The first term is manifestly divergent (as would be seen as well from 
(\ref{firstgene})), and requires some extra regularization. Since it
diverges logarithmically, 
it is natural to expect any dimensionally regularized version of this
term (such as given by the TBA) 
to go as $\ln (\hbox{cst } h^2 R)$. The next term, like all the
subsequent ones, 
is  well-defined. The sum can be evaluated in closed form: 
$$
\sum_{0}^\infty {1\over(n+1/2)(p+1/2)(n+p+1)}=7\zeta(3)
$$
giving rise to 
\begin{equation}
E_0(h)=-{\pi\over 48R}-\frac{h^2}{4\pi} \ln\left(\hbox{cst }h^2
R\right)+{\zeta(3)\over{\pi^2}} h^4 R +O(h^6 R^2).
\end{equation}

Higher orders could be obtained without too much difficulty, but the 
foregoing expression will serve our purposes. Let us now compare it
with the non perturbative results obtained from the TBA. The idea
will be recalled in detail in the next section, so for now we just give the
result, which can be found easily by taking the massless limit of 
 \cite{LeClairNPB453}:
\begin{equation}
E_0  (h)=-{1\over 4\pi R} \int_0^\infty d\kappa
\ln\left[1+{\kappa- h^2 R\over \kappa+h^2
R}~e^{-\kappa}\right].
\label{TBA} 
\end{equation}
It is not entirely straightforward to expand it in powers of $h^2R$,
in particular because the second order term 
actually involves a logarithmic piece. To proceed, if we write
$E=-{1\over 4\pi R}I(x)$, $x\equiv h^2R$, we have 
$$
\Delta I=I(x)-I(0)=\int_0^\infty d\kappa\ln\left[1-{2x\over
(1+e^\kappa)(\kappa+x)}\right].
$$
It is then  more convenient to expand the derivative as
\begin{eqnarray}
{\partial\over\partial x} \Delta I=\int_0^\infty
{d\kappa\over\kappa} {1-e^{-\kappa}\over 1+e^ 
{-\kappa}}
\sum_{n=1}^\infty (-1)^n \left({x\over\kappa}\right)^n
\left(1-e^{-\kappa}\over 1+e^{-\kappa}\right)^n +\int_0^\infty
d\kappa\left[{1\over\kappa}\left({1-e^{-\kappa}\over
1+e^{-\kappa}}\right)-{1\over \kappa+x}\right].  \nonumber 
\end{eqnarray}
Introducing the constant 
$$
c=\int_0^\infty d\kappa {1-e^{-\kappa}-2\kappa
e^{-\kappa}\over \kappa(1+\kappa)(1+e^{-\kappa})}\approx
0.125633, 
$$
one finds then, after reintegration, that
\begin{equation}
\Delta I=x(c-1)+x\ln x-\sum_{n=2}^\infty (-1)^n {(x/2)^n\over n}
\int_{-\infty}^\infty dt  \left({\tanh t\over t}\right)^n. 
\end{equation}
These integrals cannot all be evaluated in closed form, but the lowest
one can: 
$$
\int_{-\infty}^\infty dt  \left({\tanh t\over t}\right)^2={28\over
\pi^2}\zeta(3) \approx 3.41023, 
$$
from which it follows that
\begin{equation}
E(h)=-{\pi\over 48R}-\frac{h^2}{4\pi}\left(c-1+\ln h^2
R\right)+{\zeta(3)\over{\pi^2}} h^4 R +O(h^6 R^2) 
\end{equation}
This agrees  with the perturbative calculation.
In fact, one can also check the validity of the result to all orders 
(\ref{TBA}) by
comparing it with the lattice calculations  
in appendix B of \cite{YangPRB11}, so things are quite  satisfactory 
here. 

Surprises start when one considers the case of  two non-vanishing 
applied fields $h_l, h_r \neq 0$ . The TBA approach then gives 
\begin{equation}
E_0(h_l, h_r)=-{1\over 4\pi R} \int_0^\infty d\kappa
\ln\left[1+{\kappa- h_l^2 R\over \kappa+ h_l^2 R} 
~{\kappa- h_r^2 R\over \kappa+ h_r^2 R}~
e^{-\kappa}\right]\label{TBAi}
\end{equation}
This {\sl is} confusing because it {\sl seems} to have  an expansion in powers
of $h_l^2$ and $h_r^2$ only,  
a result in manifest contradiction with perturbative
calculations. Related to this puzzle is the fact that (\ref{TBAi}) is
the same 
for $h_l h_r >0$ and $h_l h_r<0$ while the two cases ought to be different
based on even simpler arguments than perturbation theory: indeed, 
in the limit of large fields, the first case describes a partition
function with fixed $++$ boundary conditions, and $E_{gs}=-{\pi\over
48 R}$,  
while the one describes a partition function with fixed $+-$ boundary
conditions, and $E_{gs}=-{\pi\over 48 R}+{\pi\over 2R}$, the
difference 
corresponding to the boundary  dimension of the spin operator,
$d={1\over 2}$.  

To make things more precise, the perturbative expansion
(\ref{firstgene}) leads, at second order, to 
\begin{equation}
E_0(h_l, h_r)=-{\pi\over 48 R}- \frac{h_l^2}{4\pi} \ln (\hbox{cst }h_l^2 R)
-\frac{h_r^2}{4\pi} R\ln
(\hbox{cst }h_r^2 R)-\frac{1}{4} h_l h_r + ...
\end{equation}
To solve the apparent contradiction with (\ref{TBAi}) we attempt to
expand it in powers of the  magnetic fields. Since we already 
have a controlled expansion for one vanishing field, e.g. $h_r = 0$,
we consider  
$$
E_0(h_l,h_r)-E_0(h_l,0)= -{1\over 4\pi R} \int_0^\infty d\kappa
\ln \left[{1 + {\kappa- h_l^2 R\over \kappa+ h_l^2
R}~{\kappa- h_r^2 R\over \kappa+ h_r^2 R}~
e^{-\kappa} 
\over 1+ {\kappa- h_l^2 R\over \kappa+ h_l^2 R}~
e^{-\kappa}}\right].
$$
This can be rewritten as 
\begin{eqnarray}
E_0(h_l, h_r)-E_0(h_l, 0)= -{1\over 4\pi R} \int_0^\infty d\kappa
\ln \left[{1 - {2 h_r^2 R\over (\kappa+ h_l^2
R)e^\kappa+\kappa- h_l^2 R}~{\kappa- h_l^2 R\over
\kappa+ h_r^2R}}\right].
\end{eqnarray}
This is not analytical in $h_r^2$, as can easily be demonstrated by
trying to calculate the derivative  
with respect to $h_r^2$ at $h_r=0$, which diverges. To proceed, we
make the change of variables $\kappa=h_l h_r Rx$ (assuming $h_l h_r
>0$),and get after some simple manipulation
\begin{eqnarray}
E_0(h_l, h_r)-E_0(h_l, 0)= -{1\over 4\pi } h_l h_r \int_0^\infty dx\left[ 1- 2~ {h_r x-h_l \over
2x+(h_r x+h_l) {e^{h_l h_r Rx}-1\over h_r}}~ { 
1\over h_l x+h_r}\right].
\end{eqnarray}
Putting $h_r =0$ in the bracket now gives a convergent integral, so 
\begin{equation}
E_0(h_l, h_r)-E_0(h_l, 0)=-\frac{h_l h_r}{4\pi} \int_0^\infty dx \ln\left[1+ {2\over
(2+h_l^2 R) x^2}\right] 
\end{equation}
and thus, the first non trivial term in $h_l h_r$ is obtained by setting
$h_l =0$ in the integral 
\begin{equation}
E_0 (h_l, h_r)-E_0(h_l, 0)=- \frac{h_l h_r}{4\pi} \int_0^\infty dx \ln\left[1+ {1\over
x^2}\right]=-\frac{1}{4} h_l h_r
\end{equation}
in agreement with the perturbative result.

We thus resolve  in a simple way the first paradox, by recognizing that
the TBA integral must 
expand in fact  powers of $h_l$ and $h_r$, not $h_{l}^{2}$,
$h_{r}^{2}$,  at least when both fields are
positive (devising a  
scheme to perform this expansion is another matter).

There does remain a bigger problem, which is that
the expansion of the TBA formula (\ref{TBAi}) is in powers of $|h_l|$
and $|h_r|$,  
and thus does not see the difference between the cases $h_l h_r >0$ and
$h_l h_r <0$ (the cases $h_l, h_r>0$ and $h_l, h_r<0$ being meanwhile 
equivalent by spin reversal symmetry). 

The reason why we have this problem  seems interesting. We do not
think
we understand it fully, but an  important hint can be found by going 
for a while to  the
initial  
description based on the ordered, massive phase (that is, before we take the massless limit). The $R$ matrix is
calculated with a single (say, right) boundary at $x=0$, but of course, the problem
is undefined until one has  
set proper boundary conditions at  (left) $x=-\infty$.  Since we are in the
ordered phase, there are two possible such conditions, spin fixed up
or down.  
For a given magnetic field on the (right) boundary, the two (left) boundary
conditions at $x=-\infty$ should correspond to different problems,
with 
presumably different  states describing the right boundary. Ghoshal and 
Zamolodchikov in \cite{GhoshalIJMPA9} argue
that spin up and $h_l >0$ correspond to the usual boundary state, while
spin down and $h_l >0$ correspond 
instead to an ``excited boundary state'', obtained formally by some
procedure of analytic continuation. We believe that  a similar feature 
holds,
the other boundary in our finite geometry 
playing the role of the boundary condition at $x=-\infty$ in \cite{GhoshalIJMPA9}.
If this is 
the case, the TBA approach proposed in \cite{LeClairNPB453} lacks a
crucial element, and is good only when, roughly speaking, the
boundary couplings would not lead to existence of frustration in the 
 integrable massive 
bulk theory  compatible with the massless boundary perturbations. In 
the Ising case: when the two fields have the same sign. 

How to handle the case $h_{l}h_{r}<0$ thus requires some additional
ingredient. To see what it may be, let us  calculate the
ground state energy of the Ising model on the strip by brute force, 
diagonalizing the hamiltonian $H$.  Following calculations initiated 
in \cite{Chatterjee}   we have
\begin{equation}
E_0 ={1\over 2} \sum_{k<0} k
\label{gschat}
\end{equation}
The modes $k$ are determined from the
quantization equation 
\begin{equation}
X\equiv e^{i2R k}~ {h_l^2+2ik \over  h_l^2-2ik}~
{h_r^2+2i k \over  h_r^2-2i k }=-1. 
\end{equation}
One can easily show that this equation has only real $k$
solutions, and that if $k$ is a solution,  
so is $-k$. Therefore, the sum in (\ref{gschat}) runs only over
the $k>0$ values. 

Now a detailed study of the quantization equation shows that all
solutions are analytical functions of $h_l^2$ and $h_r^2$ {\sl  
except} for one of them,
which in particular goes as $k=\pm k_0\approx \pm {1\over 2}|h_l h_r|$ at
small magnetic fields.  

Manipulations of (\ref{gschat}) in the usual way lead to the TBA
integral formula (\ref{TBAi}).  
Suppose now that we believe this result for $h_l h_r>0$. To get the result
for $h_l h_r<0$, we could take the formal expansion of (\ref{TBAi}) 
in powers of $h_l, h_r$ and continue analytically in the variable
$h_l h_r$. This, it seems from the previous analysis, 
involves only the modes $\pm k_0$. If the mode $+k_0$ is included
in the sum for $h_l h_r>0$, the analytically continued result 
will be obtained by discarding it, and using instead the mode
$-k_0$: in other words, we suggest that the energy for  
negative $h_l h_r$ simply reads
\begin{equation}
E_0 (h_l,h_r)=-{1\over 4\pi R} \int_0^\infty d\kappa
\ln\left[1+{\kappa- h_l^2 R\over \kappa+ h_l^2 R}~ 
{\kappa- h_r^2 R\over \kappa+ h_r^2 R} ~
e^{-\kappa}\right]+ k_0, 
\label{correctTBAi}
\end{equation}
If $h_l h_r<0$, this expression goes as $-\frac{1}{4} |h_l h_r|+ \frac{1}{2}
|h_l h_r|=\frac{1}{4}
|h_l h_r|=-\frac{1}{4} h_l h_r$, as desired. We also observe that at large  
magnetic fields, it goes to $E_0\approx -{\pi\over 48 R}+ {\pi\over
2R}$ as $k_0\rightarrow {\pi\over 2R}$, in agreement  
with the conformal prediction for $+-$ boundary conditions.

Two tests are therefore satisfied. We also observe that (\ref{correctTBAi}) 
 implies the simple result
\begin{equation}
E_0(h_l,h_r)-E_0(h_l,-h_r)=-k_0.
\label{GSEdiff}
\end{equation}
Note that the idea of adding an extra contribution in different
regimes of couplings appears first in \cite{DoreyNPB525}. 

\begin{figure}
\begin{center}
\epsfig{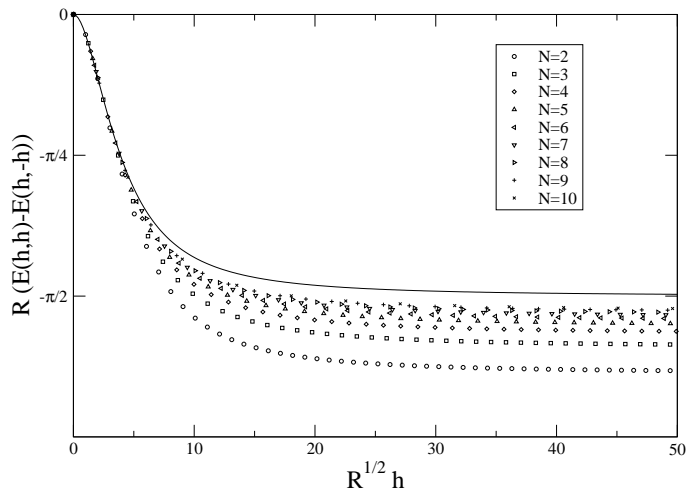}
\vspace{0.7cm}
\caption{Ground-state energy difference for the two-boundary Ising
model with boundary magnetic fields $h_l = h_r$ and $h_l = - h_r$.
This is plotted as a function of the scaling parameter $R^{1/2} h$
for system sizes up to ten lattice sites,
and compared to the theoretical prediction in equation
(\ref{GSEdiff}), i.e. the root $k_0$.}
\label{fig:IsingGSEdiff}
\end{center}
\end{figure}

To provide supplementary justification for this construction, we have
performed some numerical checks on equation (\ref{GSEdiff}) using
Mathematica.  The plots were produced using exact diagonalization of 
a lattice Ising model at its bulk critical point,  for
system sizes up to ten sites, with equal absolute values for the left
and right boundary magnetic fields.  Both cases $h_l = h_r$ and $h_l =
- h_r$ were separately solved, and their difference compared to the
root $k_0$.  Figure \ref{fig:IsingGSEdiff} shows our results.
Difficulties related with the relative values of the two boundary
parameters will be generic - and worse - in the case of the double
boundary sine-Gordon model, and require considerable efforts. On top 
of this, an additional difficulty arises due to the presence of the
phase $\chi$: even if the bulk is chosen to be at the reflectionless 
points and hence have diagonal scattering, the boundary reflections
cannot be simultaneously diagonalized, requiring substantial
modification to the crossed channel TBA of \cite{LeClairNPB453}. We
would like to illustrate this in the case of the free fermion point
$\beta^{2}=4\pi$. 

\section{All the difficulties in a nutshell: the double boundary sine-Gordon model at  
$\beta^{2}=4\pi$.}

\subsection{The crossed channel TBA}

At the free fermion point $\beta^2 = 4\pi$, the  (massive) sine-Gordon theory
simplifies considerably.  The only fundamental excitations are two fermions, 
the soliton and the antisoliton.  Their
scattering matrix in the bulk is trivial, independent of
rapidity - the creation and annihilation operators obey the canonical
anticommutation relations
\begin{eqnarray}
\{ A_a (\theta), A^{\dagger}_b (\theta ') \} = \delta_{ab}
\delta(\theta - \theta ')
\end{eqnarray}
with all other anticommutators vanishing, $a=\pm$ labelling soliton/antisoliton particle types.  
Scattering at the boundaries is encoded in the $R$ matrix
\begin{equation}
    \left(\begin{array}{cc}
    R_{++}&R_{+-}\\
    R_{-+}&R_{--}\end{array}\right)\equiv\left(\begin{array}{cc}
    P_{+}&Q\\
    Q&P_{-}\end{array}\right)
\end{equation}
There will be one reflection matrix for each boundary, and its
elements will depend on the rapidity of the incident particle, as
well as the bulk and boundary parameters of the field theory. 
It is easier to obtain the reflection amplitudes formally
 by solving the boundary bootstrap equations; one is thus left with
 some undetermined relation between the parameters entering the $R$
 matrix, and the parameters in the field theory. We will have to
 tackle this issue in more details later. For now, we observe (as
 discussed in details in the appendix) that the point
 $\beta^{2}=4\pi$ being equivalent to a free fermion theory,  explicit
calculations can easily be carried out, and one finds \cite{AmeduriPLB354} 
\begin{eqnarray}
P^{\pm}_{l,r} (\theta) &=& [\cosh \theta - \frac{1}{2} \gamma_{l,r} \cosh(\theta 
\mp i\phi_{l,r})] /D_{l,r} (\theta), \nonumber \\
Q^+_{l,r} (\theta) &=& Q^-_{l,r} (\theta) = \frac{i}{2} \frac{\sinh 
2\theta}{D_{l,r}(\theta)},  \\
D_{l,r} (\theta) &=& -i \gamma_{l,r} \cosh (\frac{\theta - i 
\phi_{l,r} - i\pi/2}{2})   
\sinh (\frac{\theta +i\phi_{l,r} +i\pi/2}{2}) - \cosh^2 \theta, 
\end{eqnarray}
where $\gamma_{l,r}  \propto \tilde{\Delta}_{l,r}^2/m \nonumber$ and $\phi_l =0, \phi_r = \phi_0$.  

In the $R$-channel description of the problem, the explicit form of
the boundary states then reads
\begin{eqnarray}
|B_{l,r}> \propto \exp \left( \int_{0}^{\infty} d\theta
K^{ab}_{l,r} (\theta) A^{\dagger}_a(-\theta) A^{\dagger}_b(\theta)
\right) |0>.  
\label{boundarystates}
\end{eqnarray}
Here, the $K$ matrices are given by the analytic continuation of the 
boundary R-matrix according to\begin{eqnarray}
K_{l,r} (\theta) &=& \frac{i/2}{\sinh^2 \theta + \gamma_{l,r}
\cosh\left(\frac{\theta +i\phi_{l,r}}{2}\right) \cosh\left(
\frac{\theta -i\phi_{l,r}}{2}\right)} \times \nonumber \\
&& \times 
\left( \begin{array}{cc}
\sinh 2\theta & -2 \sinh \theta + \gamma_{l,r} 
\sinh(\theta-i\phi_{l,r}) \\  
 -2 \sinh \theta + \gamma_{l,r} \sinh(\theta+i\phi_{l,r}) & \sinh 
 2\theta   
\end{array} \right).
\label{K}
\end{eqnarray}
As it should, this obeys the boundary cross-unitarity condition $K^{ab} 
(\theta) = - K^{ba} (-\theta)$, in view of the fact that the scattering 
matrix is simply $-1$.

It is now straightforward to take the massless limit 
through
\begin{eqnarray}
m \rightarrow 0, \hspace{1cm} \theta_0 \rightarrow \infty, \hspace{1cm} 
me^{\theta} = m e^{\theta_0 + \theta'} = \mu e^{\theta'}, 
\hspace{1cm} \mu \hspace{0.2cm} finite.
\end{eqnarray}
$\theta'$ (which we will often denote by $\theta$ when no ambiguity is possible) 
is then integrated from $-\infty$ to $\infty$. We will often denote 
$\mu e^{\theta'}\equiv\kappa$, and in practice set $\mu=1$ (sometimes 
the massless limit is taken with $\mu=2$ instead, which does not affect 
any of the results. Also, we use the notation $\mu$ briefly in appendix C, but no confusion should be possible). Performing the 
massless limit on the $K$-matrix (\ref{K}), we get
\begin{eqnarray}
K_{l,r}(\theta') = \frac{i}{\mu e^{\theta'} + \Delta^2_{l,r}}
\left( \begin{array}{cc}
\mu e^{\theta'} & \Delta_{l,r}^2 e^{-i\phi_{l,r}} \\
\Delta_{l,r}^2 e^{i\phi_{l,r}} & \mu e^{\theta'} 
\end{array} \right).\label{masslesskff}
\end{eqnarray}
(Note that some care has to be exercised in getting the $K$ matrix in 
the massless limit directly from the $R$ matrix in the same limit). The couplings $\Delta$ are renormalized couplings, in this case $\Delta=\sqrt{\pi}\tilde{\Delta}$. 

Equipped with this data, we now consider the calculation of the
ground state energy generalizing \cite{LeClairNPB453}.  Although this
was already carried out in \cite{CauxPRL88}, we include here some more
extensive explanations, which will be useful in the following sections.

Let us now turn to the calculation of the partition function in the limit 
$L \rightarrow \infty$, where the boundary states are well-defined.  Our
starting point is a version of equation (\ref{RchannelZ}):
\begin{eqnarray}
Z = \sum_{\alpha} \frac{\langle B_l | \alpha \rangle \langle \alpha | 
B_r \rangle}{\langle \alpha | \alpha \rangle} e^{-R E_{\alpha}},
\label{RchannelZagain}
\end{eqnarray}
where $\alpha$ is an index running through all states with nonzero inner
product with the boundary states.  In view of the structure of the boundary 
states, we will thus need to trace over all states of the form
\begin{eqnarray}
| \alpha \rangle \in \{  ~A^{\dagger}_{a_N} (-\theta_N) A^{\dagger}_{b_N} 
(\theta_N) ... A^{\dagger}_{a_1} (-\theta_1) A^{\dagger}_{b_1} 
(\theta_1) |0 \rangle, ~\forall ~N ~\in ~{\Bbb N}, ~\forall ~\theta ~\in 
~{\Bbb R} \}.
\label{states}
\end{eqnarray}
We have to be very careful here:  in the case of a theory with only one 
type of particle, the rapidities can be ordered using the strict 
inequality $\theta_N > ... > \theta_1$, since the Bethe wave functions
vanish identically for coinciding rapidities (see e.g. \cite{LeClairNPB453}).  
Now, however, we have more than one species of particles, so we ${\it can}$
have coinciding rapidities without having vanishing Bethe wave functions, 
as long as the labels of particles at coinciding rapidities do not get
reproduced.  A more economical strategy consists in introducing creation
operators for particle ${\it pairs}$, and a ${\it quartet}$ one 
\cite{CauxPRL88}:
\begin{eqnarray}
P^{\dagger}_{ab} (\theta) \equiv A^{\dagger}_a(-\theta) A^{\dagger}_b 
(\theta), \hspace{0.5cm} 
Q^{\dagger} (\theta) \equiv A^{\dagger}_1(-\theta) A^{\dagger}_1(\theta)
A^{\dagger}_2(-\theta) A^{\dagger}_2(\theta), \hspace{0.5cm} 
\theta > 0
\label{pairandquartet}
\end{eqnarray}
where the pair creation operators come in four flavours, labeled 
$11, 22, 12$ and $21$.
We can now consistently restrict the rapidities to be positive and
noncoincident in the set of states $| \alpha \rangle$.  Note that 
because of basic anticommutation relations and the Pauli principle, the 
structure of the boundary state dictates that the quartet appears with 
fugacity $K^{11}(\theta) K^{22}(\theta) - K^{12}(\theta) K^{21}(\theta)
= \det K$.

We can now rewrite our trace over states in (\ref{RchannelZagain}) as 
running over the set of all states of the form
\begin{eqnarray}
| \alpha \rangle \in \{ ~P^{\dagger}_{a_N b_N}(\theta_N) ... 
P^{\dagger}_{a_1 b_1} (\theta_1) Q^{\dagger} (\theta^q_{N_q}) ... 
Q^{\dagger} (\theta^q_1) |0 \rangle,  \nonumber \\
~\forall ~N,N_q ~\in ~{\Bbb N}, 
~\theta_N > ... > \theta_1, ~\theta^q_{N_q} > ... > \theta^q_1 ,
~\theta_i \neq \theta^q_j \}.
\end{eqnarray}
For a state with $N$ pairs with rapidities $\theta_1,... ,\theta_N$
and $N_q$ quartets with rapidities $\theta^q_1,... ,\theta^q_{N_q}$, 
we can easily show by explicitly expanding the exponentials in the
expressions for the boundary states that
\begin{eqnarray}
\frac{\langle B_l | \alpha \rangle \langle \alpha | 
B_r \rangle}{\langle \alpha | \alpha \rangle} = \prod_{i=1}^N 
\bar{K}^{b_i a_i}_l (\theta_i) K^{a_i b_i}_r (\theta_i) 
\prod_{j=1}^{N_q} \det [\bar{K}_l (\theta^q_j) K_r(\theta^q_j)]
\end{eqnarray}
To construct the entropy of a field configuration, necessary to complete the
TBA, we use the following argument.  Imagine putting our system in a very 
large but finite domain of length $L$ in the space-like direction of the
R-channel picture.  Since all bulk scattering is given by the trivial 
scattering matrix $S = -1$, the quantization rules dictate that the
allowed rapidities appearing in the pair and quartet creation operators
lie on a lattice of almost evenly spaced values (separated by intervals 
of order $1/L$, and symmetric with respect to sign exchange).  Moreover,
in view of the Pauli principle, no superpositions of creation operators
can take place on a given rapidity.  Thus,
the entropy of a state configuration with $N_{ab}$ pairs of type $a,b$ and
of $N_q$ quartets is very simple to calculate, being that of a 
gas of mutually excluding particles of five different types.
Explicitly, for $N_T$ total allowed rapidities, we
have the simple formula
\begin{eqnarray}
e^S = \frac{N_T !}{N_{11}! N_{22}! N_{12}! N_{21}! N_q ! (N_T - N_{11} - 
N_{22} -N_{12} -N_{21} -N_q )!}.
\end{eqnarray}
Introducing the usual occupation densities $\rho, \rho_{ab}, \rho_q$ as well
as the hole density $\rho_h = \rho - \sum_{ab} \rho_{ab} - \rho_q$
as we take the limite $L \rightarrow \infty$, we can
rewrite this in the Stirling approximation as
\begin{eqnarray}
S \approx L \int_0^{\infty} d \theta \left[ \rho \ln \rho - \sum_{ab} 
\rho_{ab} \ln \rho_{ab} - \rho_q \ln \rho_q - \rho_h \ln \rho_h \right].
\end{eqnarray}
The energy is on the other hand given by
\begin{eqnarray}
E = L \int_0^{\infty} d \theta \left[ \sum_{ab} \rho_{ab} + 2 \rho_q 
\right] 2m \cosh \theta,
\end{eqnarray}
allowing us to put everything together to obtain the partition function
\begin{eqnarray}
Z \approx \int [d\rho] \exp \biggl\{ L \int_0^{\infty} d \theta \biggl[
\sum_{ab} \left(\ln \bar{K}^{ba}_l K^{ab}_r - 2mR \cosh \theta \right)
\rho_{ab} + \left( \ln \det [\bar{K}_l K_r] - 4mR \cosh \theta
\right) \rho_q + S[\rho] \biggr] \biggr\}.
\end{eqnarray}
Introducing the quasienergies $e^{-\epsilon_{ab}} = \rho_{ab}/\rho_h$ and 
a similar one for the quartets, we can obtain the TBA equations by using a 
saddle-point approximation on the partition function.  From the
quantization equation we get the constraint $\rho_h  (1+
\sum_{ab} e^{-\epsilon_{ab}} + e^{-\epsilon_q}) = \frac{m}{2\pi} \cosh
\theta$, which when substituted in the saddle-point equation leads 
(after some straightforward manipulations) to
\begin{eqnarray}
\ln Z &=& \frac{m L}{2\pi} \int_0^{\infty} d \theta \cosh \theta 
\ln \left( 1 + \sum_{ab} e^{-\epsilon_{ab}} + e^{-\epsilon_q} \right),
\nonumber \\
\epsilon_{ab} &=& 2mR \cosh \theta - \ln \bar{K}^{ba}_l K^{ab}_r, \nonumber \\
\epsilon_q &=& 4mR \cosh \theta - \ln \det(\bar{K}_l K_r).
\end{eqnarray}
Reinterpreting this as the ground-state energy of the original problem,
we arrive at the final formula
\begin{equation}
E_0 = -\frac{1}{2\pi} \int_0^{\infty} d \theta m \cosh \theta ~\ln \biggl[
1 + tr (\bar{K}_l K_r) e^{-2mR \cosh \theta}  + \det (\bar{K}_l K_r)
e^{-4mR \cosh \theta} \biggr].
\end{equation}
\label{TBAgroundstate}
where the $K$ matrices are given in equation (\ref{K}).

We can now take the massless limit using the results
(\ref{masslesskff}) to obtain the final expression
\begin{eqnarray}
E_0 = -\frac{1}{4\pi} \int_0^{\infty} d \kappa \ln \left[ 1+2
\frac{\kappa^2 + \Delta_l^2 
\Delta_r^2 \cos 2\chi}{(\kappa +  \Delta_l^2)(\kappa +  \Delta_r^2)}
e^{-\kappa R} 
+ 
\frac{(\kappa -  \Delta_l^2)(\kappa -  \Delta_r^2)}{(\kappa +  \Delta_l^2)(\kappa
+  \Delta_r^2)} e^{-2\kappa R}
\right] 
\label{groundstate}
\end{eqnarray}
In the first appendix, we provide the details of an independent derivation  of this
result using refermionization, confirming that the TBA construction that we have
employed is justified.

\subsection{Analytic structure of the ground-state energy}

The study of the analytical properties of the ground-state energy of
the double-boundary Ising model in the first part of this paper
has shown how careful one has to be in
interpreting the data coming from the TBA equations.   It is here
again very easy to miss out on the correct analytic continuation
procedure to be imposed.  

Let us thus carefully study our expression for the ground-state energy
of the double-boundary sine-Gordon model at the free fermion point,
equation (\ref{groundstate}), whose bulk term we reproduce here:
\begin{eqnarray}
E_0 = -\frac{1}{4\pi} \int_0^{\infty} d \kappa \ln \left[ 1+2
\frac{\kappa^2 + \Delta_l^2 
\Delta_r^2 \cos 2\chi}{(\kappa +  \Delta_l^2)(\kappa +  \Delta_r^2)}
e^{-\kappa R} 
+ \frac{(\kappa -  \Delta_l^2)(\kappa -  \Delta_r^2)}{(\kappa + 
\Delta_l^2)(\kappa 
+  \Delta_r^2)} e^{-2\kappa R}
\right].
\label{bulkgroundstate}
\end{eqnarray}

First of all, we expect the ground state energy to be an even function
of the phase $\chi$, as is obvious from the boundary sine-Gordon Hamiltonian.  
We can consequently concentrate on positive $\chi$ from now on.  The
first crucial observation is that equation (\ref{bulkgroundstate}) is
valid only in the domain $\chi \in [0, \pi/2]$, since it exhibits a
singularity at $\chi = \pi/2$:  for $\chi \in [0, \pi/2[$, the
argument of the logarithm has zeroes in the complex plane of the
variable $\kappa$ that hit the real axis at $\kappa = 0$ when $\chi$
reaches $\pi/2$.

We can gain a bit more intuition about this formula by considering the
limits $\Delta_l, \Delta_r \rightarrow \infty$, where the integral reduces to
\begin{eqnarray}
E_0 = -\frac{1}{4\pi R} \int_0^{\infty} dx \ln \left[ 1 + 2 \cos 2\chi
e^{-x} + e^{-2x} \right].
\end{eqnarray}
This integral is tabulated, and we get
\begin{eqnarray}
E_0 = - \frac{\pi}{24 R} + \frac{\chi^2}{2 \pi R}, \hspace{1cm} \chi
\in [0, \pi/2], \nonumber \\
E_0 = - \frac{\pi}{24 R} + \frac{(\chi - \pi)^2}{2 \pi R}, \hspace{1cm}
\chi \in [\pi/2, \pi].
\end{eqnarray}
As a function of the phase difference $\chi$, the basic integral
representation (\ref{groundstate}) for the ground-state energy thus
has a cusp at $\chi = \pi/2$.  

A simple computation shows that this should not be the case.  Namely,
let us consider the boundary Hamiltonian (\ref{boundaryH}) in the limit
$\Delta_{l,r} \rightarrow \infty$.  In that case, the field becomes
pinned at $x=0$ to $\phi (0,t) = 2\sqrt{\pi} (n + 1/2)$, while at
$x=R$ it gets pinned to $2\sqrt{\pi} (m + 1/2) + 
\chi/\sqrt{\pi}$, where $n,m \in {\Bbb Z}$.  This immediately tells us
that the zero mode operator $\Pi_0$ has to have eigenvalue $\Pi_0 = 2
\sqrt{\pi} (p - \frac{\chi}{2\pi}), p \in {\Bbb Z}$.  For the phase
$\chi$ in the fundamental domain $|\chi| < \pi$, the lowest-energy
state is thus the one with $p=0$, so the energy simply becomes (using
 (\ref{Hamiltonianmodes}) and adding specifically 
the contribution from the dynamical modes)
\begin{eqnarray}
E_0 = -\frac{\pi}{24R} + \frac{\chi^2}{2\pi R}, \hspace{2cm} |\chi| <
\pi
\end{eqnarray}
exactly.  This thus clearly demonstrates that the basic integral
representation (\ref{groundstate}) works in the domain $|\chi| <
\pi/2$, but fails for $\pi/2 < |\chi| < \pi$.  

Let us now study the case of
small $\Delta_{l,r}$.  From the integral representation for the
ground-state energy (\ref{groundstate}), we can write (after factorizing
$(1+e^{-\kappa})^2$ in the logarithm and a few more basic
manipulations) 
\begin{eqnarray}
E_0 = -\frac{\pi}{24R} -\frac{1}{4\pi R} \int_0^{\infty} d \kappa \ln
\left[ 1-\frac{4\eta (\kappa (1+e^{-\kappa}) + \eta \sin^2 \chi)
e^{-\kappa}}{(1+e^{-\kappa})^2 (\kappa + \eta)^2} \right]
\end{eqnarray}
where we have for simplicity considered $\Delta_l = \Delta_r$ and
defined the dimensionless parameter $\eta = \Delta_{l,r}^2 R$.  One
sees that it is not possible to do a simple Taylor expansion in $\eta$
in this formula.  Rather, to evaluate it in this limit, we observe
that the main contributions come from the region of integration where
$\kappa \sim \eta$.  Defining $\xi$ such that $\eta << \xi << 1$, we
can approximate the integral term as
\begin{eqnarray}
I \approx - \frac{1}{4\pi R} \int_0^{\xi} d \kappa \ln \left
[ 1-\frac{\eta(2\kappa + \eta \sin^2 \chi)}{(\kappa + \eta)^2} \right]
= -\frac{1}{4\pi R} \int_0^{\xi} d \kappa \ln \left[ \frac{\kappa^2 +
\eta^2 \cos^2 \chi}{(\kappa + \eta)^2} \right].
\end{eqnarray}
This can in turn be approximated by rescaling by $\eta$ and replacing
the upper bound on the integral by infinity.  The resulting integral
is tabulated, and we finally obtain
\begin{eqnarray}
E_0 \approx -\frac{\Delta_{l,r}^2}{4} |\cos \chi| + cst.
\end{eqnarray}
The absolute value gives rise to the same kind of cusp as the one 
obtained in the limit $\Delta_{l,r} \rightarrow \infty$.  Here,
we can compare with straightforward perturbation theory which gives us
$\cos \chi$ instead of its absolute value.  

These considerations again indicate the presence of a 
problem with the integral representation (\ref{bulkgroundstate})
around the point $\chi = \pi/2$.  In order to repair this, we can first use
a procedure very similar to the one used in the Ising case.  Namely,
we identify the ground-state energy $E_0$ with the integral
representation (\ref{bulkgroundstate}) only in the domain $|\chi| <
\pi/2$.  The domain $\pi/2 < |\chi| < \pi$ is accessed by analytic
continuation, which is performed based on the explicit diagonalization (see appendix A).  First, note that all the
roots of the quantization condition (\ref{quantization}) are
well-behaved as a function of $\chi$ except for the pair closest to
the origin, which we label as $ \pm k_0$.  This is the root that we
need to analytically continue, as in the Ising case.  Thus, in the
domain $\pi/2 < |\chi| < \pi$, the true ground-state energy should be
given by the integral representation (\ref{bulkgroundstate}) plus
$k_0$. 

The validity of this procedure can again be checked in the limits of
strong and weak boundary pairings.  In the limit $\Delta_{l,r}
\rightarrow \infty$, the root $k_0$ is given by $k_0 = \frac{\chi -
\pi/2}{R}$.  This leads, in the domain $\pi/2 < |\chi| < \pi$, to a
ground-state energy of $E_0 = \frac{-\pi}{24 R} + \frac{(\chi -
\pi)^2}{2 \pi R} + \frac{\chi - \pi/2}{R} = \frac{-\pi}{24 R} +
\frac{\chi^2}{2\pi R}$, which is the result required by the above
considerations using the zero-mode operators.  In the weak boundary
pairing limit $\Delta_{l,r} \rightarrow 0$, the root $k_0$ behaves as
$-{1\over 2}\Delta_{l,r}^2 \cos \chi$, leading in the domain $\pi/2 <|\chi| <
\pi$ to $E_0 = -\frac{\Delta_{l,r}^2}{4} |\cos \chi| - {1\over 2}\Delta_{l,r}^2
\cos \chi + cst. = -\frac{\Delta_{l,r}^2}{4} \cos \chi + cst.$, as
required by simple perturbation theory.

To summarize, we write our final result as
\begin{eqnarray}
E_0^{SG} (\Delta_l, \Delta_r, \chi) = E_0 (\Delta_l, \Delta_r, \chi),
\hspace{3cm} |\chi| \leq \pi/2, \nonumber \\
E_0^{SG} (\Delta_l, \Delta_r, \chi) = E_0 (\Delta_l, \Delta_r, \chi) +
k_0, \hspace{2cm} \pi/2 \leq |\chi| < \pi,
\label{finalGSEatFFP}
\end{eqnarray}
with $2\pi$ periodicity in $\chi$, and where $E_0$ is the integral
(\ref{bulkgroundstate}). 
Plots of this ground-state energy as a function of the phase difference
$\chi$ can be found in \cite{CauxPRL88} for various values of the
boundary parameters.

To extend the method away from the free fermions point, it is necessary first to 
reinterpret the addition of $k_0$ directly within the TBA approach; this will help
make the case $\lambda = 2$ more transparent. We start by writing
down again our results from the TBA.  For $\lambda = 1$, since the
scattering in the bulk is trivial, the quasienergies obey
the trivial equation 
\begin{eqnarray}
\epsilon(\kappa) = \kappa
\label{simpleTBA}
\end{eqnarray}
where the ground-state energy has the integral representation 
(we have rescaled the boundary parameters according to $\Delta_{L,R} = R^{1/2}
\Delta_{l,r}$)
\begin{eqnarray}
E_0 = -\frac{1}{4\pi R} \int_0^{\infty} d \kappa \ln \left[ 1+2
\frac{\kappa^2 + \Delta_L^2 
\Delta_R^2 \cos 2\chi}{(\kappa +  \Delta_L^2)(\kappa +  \Delta_R^2)}
e^{-\epsilon(\kappa)} 
+ \frac{(\kappa -  \Delta_L^2)(\kappa -  \Delta_R^2)}{(\kappa + 
\Delta_L^2)(\kappa 
+  \Delta_R^2)} e^{-2\epsilon(\kappa)}
\right].
\label{simpleGSE}
\end{eqnarray}
and we have performed some simple rescaling. 
Now of course, the equation (\ref{simpleTBA}) defines the quasienergy everywhere
in the complex plane of $\kappa$.  In particular, $\epsilon (\kappa)$ is
pure imaginary if and only if $\kappa$ is also pure imaginary.  Moreover,
this equation does not need to be modified as a function of $\chi$, i.e. there
is no analytical continuation needed for the quasienergy.

If, however, we closely examine the ground-state energy, it is clear that
something goes wrong.  Looking at the integrand at $\kappa = 0$ yields
\begin{eqnarray}
\ln \left[ 1 + 2 \cos 2\chi e^{-\epsilon(0)} + e^{-2\epsilon(0)} \right]
= 2 \ln \cos \chi
\end{eqnarray}
which clearly diverges when $\chi \rightarrow \pi/2$, so we know that 
there is a singularity hitting $\kappa = 0$ at $\chi = \pi/2$. 
For general $\chi$, the position of this singularity is found by selecting
the proper solution of the singularity condition
\begin{eqnarray}
e^{2 \epsilon(\kappa)} + 2 \frac{\kappa^2 + \Delta_L^2 \Delta_R^2 
\cos 2\chi}{(\kappa + \Delta_L^2)(\kappa + \Delta_R^2)} e^{\epsilon(\kappa)}
+ \frac{(\kappa - \Delta_L^2)(\kappa - \Delta_R^2)}
{(\kappa + \Delta_L^2)(\kappa + \Delta_R^2)} = 0.
\end{eqnarray}
A careful look at this shows that we must look for such a singularity
at values of $\kappa$ for which $\epsilon(\kappa)$ is purely imaginary,
and therefore, according to what we wrote before, on the imaginary axis of
$\kappa$.  We therefore note the singularity position as $i \kappa_s$
(note that if $i\kappa_s$ is a singularity, so is $-i\kappa_s$).  The
picture is then as follows:  singularities at $\pm i \kappa_s$ touch
zero at $\chi = \pi/2$, and the ground-state energy integral 
(\ref{simpleGSE}) must be modified to take this into account.  Simple
contour-integral manipulations then allow to write the modification
to the ground-state energy, which should be added for $\chi > \pi/2$, 
in the form $
\delta E_0 = \frac{\kappa_s}{2R}$
making contact with the previous mode expansion arguments.  

This 
is how the analytical continuation should be understood for the case
$\lambda = 1$ in the language of the TBA, and the logic of this
procedure will allow us to find the solution to the analytic 
continuation problem in the case $\lambda = 2$.  There, things will be
more complicated:  in particular, the self-consistency equations will 
themselves have to be analytically continued, and the singularity
positions will lie in a different domain of the complex plane.

\section{The double boundary sine-Gordon model: some general results at the reflectionless points.}

\subsection{The ground state energy}

 The massive sine-Gordon theory has its 
discrete symmetry $\phi \rightarrow \phi + \frac{2\pi}{N}\beta, N \in
{\Bbb N}$  spontaneously broken for $\beta^2 < 8\pi$.  The spectrum 
has a number of fundamental particles
classified as a soliton-antisoliton pair, and  breathers
(bound states).
Although the soliton and the antisoliton are always present, the
number of breathers is dictated by the value of the interaction
parameter $\beta$.  Namely, the breathers are labeled with the integer
$n$ s.t. $n=1,2,...<\lambda$ where
\begin{equation}
\lambda = \frac{8\pi}{\beta^2} -1.
\end{equation}
Solitons and antisolitons carry respectively positive and negative
topological charge, whereas the breathers are neutral.  If the soliton
mass is given by $m$, then the breathers carry masses
\begin{equation}
m_n = 2m \sin \frac{n\pi}{2\lambda}, n=1,2,...<\lambda.
\end{equation}
Let us adopt like in the free fermion case the notation $A^{\dagger}_+ (\theta)$ and
$A^{\dagger}_- (\theta)$ for the creation operators at rapidity
$\theta$ of the soliton and
antisoliton, respectively.  The creation operators for the $n$-th
breather at rapidity $\theta$ will be denoted $B^{\dagger}_n (\theta)$.
Factorized scattering in terms of these is described by the fundamental
scattering amplitudes
\begin{eqnarray}
A_{\pm}^{\dagger}(\theta_1) A_{\pm}^{\dagger} (\theta_2) &=&
 a(\theta_1 - \theta_2) 
 A_{\pm}^{\dagger}(\theta_2) A_{\pm}^{\dagger} (\theta_1),  \nonumber \\
A_+^{\dagger}(\theta_1) A_-^{\dagger} (\theta_2) &=& b(\theta_1 - \theta_2)
A_-^{\dagger}(\theta_2) A_+^{\dagger} (\theta_1) + c(\theta_1 -
 \theta_2) A_+^{\dagger}(\theta_2) A_-^{\dagger} (\theta_1).
\label{amplitudes}
\end{eqnarray}
The amplitudes for these scattering processes are well-known 
\cite{ZamolodchikovAP120}, and take the form either of infinite
products of gamma functions or of more compact integral
representations.  These are reproduced in appendix B.

The introduction of a boundary interaction is described like in the
free fermion case by a reflection matrix on the left and right
boundaries respectively. In addition, there is an amplitude
$B_{n}(\theta)$ for each breather - the scattering cannot mix
particles, except solitons and antisolitons.

The boundary boostrap method applied to this case yields the four
amplitudes $P_{\pm}, Q_{\pm}$ in terms of two free parameters $\xi, k$
which we will call ``IR'' parameters following \cite{Takacs}.
Their relationship to the ``UV'' parameters $\Delta, \chi$ appearing
in the boundary Lagrangian is a complicated issue in general, which we
will address later on.  Suffice to say now that the results of the
boundary bootstrap for the boundary scattering amplitudes are
summarized in appendix C.  In particular, we have $\beta\phi_{0}/2\equiv \chi$. 

We will restrict in what follows to 
 reflectionless points where the bulk scattering is diagonal (that is, $c(\theta) = 0$).  They occur at
$\lambda \in {\Bbb N}$.  In these cases, there will be $\lambda - 1$
breathers in the theory, with the original interaction 
parameter given by $\beta^2 = \frac{8\pi}{\lambda + 1}$.
As in the free fermion case, the solitons and antisolitons have to be
carefully treated when ordering the rapidities.  As detailed before,
it is better to move to a basis where only positive rapidites are
used, and the states involved in the trace for the partition function 
are constructed using pair and quartet creation operators
$P_{ab}^{\dagger}, Q^{\dagger}$ (see the discussion around equation
(\ref{pairandquartet})).  Here, we will need to add the breather pair
creation operator, defined as
\begin{eqnarray}
R^{\dagger}_{n}(\theta) \equiv B^{\dagger}_n(-\theta)
B^{\dagger}_n(\theta), \hspace{2cm} \theta > 0. 
\end{eqnarray}
Now, we simply need to add all the possible breather pair 
states to our set of states to trace over:
\begin{eqnarray}
| \alpha \rangle \in \{ ~P^{\dagger}_{a_N b_N}(\theta_N) ... 
P^{\dagger}_{a_1 b_1} (\theta_1) Q^{\dagger} (\theta^q_{N_q}) ... 
Q^{\dagger} (\theta^q_1) 
R^{\dagger}_{n_{N_b}}(\theta^b_{N_b})... R^{\dagger}_{n_1}(\theta^b_1)
|0 \rangle,  \nonumber \\
~\forall ~N,N_q, N_b ~\in ~{\Bbb N}, 
~\theta_N > ... > \theta_1, ~\theta^q_{N_q} > ... > \theta^q_1 ,
~\theta_i \neq \theta^q_j \}.
\end{eqnarray}

In order to compute the partition function in the thermodynamic limit,
let us start with the quantization conditions for the allowed
rapidities of the different types of excitations.  For definiteness,
let us consider a state with $N$ pairs with rapidities
$\theta_1,...,\theta_N$, $N_q$ quartets with rapidities
$\theta^q_1,...,\theta^q_{N_q}$ and $N_b$ breather pairs with
rapidities $\theta^b_1,...,\theta^b_{N_b}$.  Let us imagine moving the
particle in pair $i$ having rapidity $\theta_i$ around the system, and
requiring periodicity of the wavefunction.  
This leads to the
quantization condition
\begin{eqnarray}
e^{i m L \sinh \theta_i} S(2 \theta_i) \prod_{j (\neq i) =1}^N
S(\theta_i - \theta_j) S(\theta_i + \theta_j) 
\prod_{k=1}^{N_q} S^2(\theta_i - \theta^q_k) S^2(\theta_i +
\theta^q_k)
\prod_{l=1}^{N_b} S^{(n_l)}(\theta_i - \theta^b_l) S^{(n_l)}(\theta_i
+ \theta^b_l) = 1.
\label{quantization1}
\end{eqnarray}
The other quantization condition is obtained by moving one member of
the breather pairs around the system in a similar fashion:
\begin{eqnarray}
e^{i m_n \sinh \theta^b_i} S^{(n_i, n_i)} (2 \theta^b_i) 
\prod_{j=1}^N S^{(n_i)}(\theta^b_i - \theta_j) S^{(n_i)}(\theta^b_i +
\theta_j) \times \nonumber \\
\times 
\prod_{k=1}^{N_q} {S^{(n_i)}}^2 (\theta^b_i - \theta^q_k)
{S^{(n_i)}}^2 (\theta^b_i + \theta^q_k) 
\prod_{l (\neq i)=1}^{N_b} S^{(n_i, n_l)} (\theta^b_i - \theta^b_l)
S^{(n_i, n_l)} (\theta^b_i + \theta^b_l) =1. 
\label{quantization2}
\end{eqnarray}
As usual, we define densitites of occupied and unoccupied rapidities
but now for pairs, quartets and breather pairs.  A complete set of
states is obtained by considering positive rapidities only (remember:
our pairs have one component at positive rapidity, and one at negative
rapidity).  For example, $[\rho_{ab} +\rho_{ab}^h] L d \theta$ gives
the number of allowed rapidities in the interval $\theta, \theta +
\delta \theta$ for pairs of type $ab$, with $\rho_{ab}$ the density of
filled states.  The symmetry properties $\rho_{ab} (-\theta) =
\rho_{ab} (\theta), \rho_q (-\theta) = \rho_q (\theta)$ and $\rho_n
(-\theta) = \rho_n (\theta)$ can then be used to simplify the
resulting equations.  In logarithmic form, we get (using the usual notation
$a*b (\theta) = \int_{-\infty}^{\infty} d\theta' a(\theta - \theta') b(\theta')$)
\begin{eqnarray}
\sum_{ab} \rho_{ab} (\theta) + \rho_q (\theta) + \rho^h (\theta) = 
\frac{m}{2\pi} \cosh \theta - \Phi * \left[ \sum_{ab}
\rho_{ab} (\theta) + 2 \rho_q (\theta) \right]
- \sum_{n=1}^{\lambda-1} \Phi^{(n)} * \rho_n
(\theta),
\nonumber \\
\rho_n (\theta) + \rho_n^h (\theta) = \frac{m_n}{2\pi} \cosh \theta -
\Phi^{(n)} * \left[ \sum_{ab}
\rho_{ab} (\theta) + 2 \rho_q (\theta) \right]
- \sum_{m=1}^{\lambda-1} \Phi^{(n,m)} * \rho_m
(\theta),
\label{RLquantizationconditions}
\end{eqnarray}
where the integration kernels are obtained from the scattering
matrices according to
\begin{eqnarray}
\Phi (\theta) &=& \frac{-1}{2\pi i} \frac{d}{d \theta} \ln S(\theta),
\nonumber \\
\Phi^{(n)} (\theta) &=& \frac{-1}{2\pi i} \frac{d}{d \theta} \ln S^{(n)}
(\theta), \nonumber \\
\Phi^{(n,m)} (\theta) &=& \frac{-1}{2\pi i} \frac{d}{d \theta} \ln
S^{(n,m)} (\theta),
\end{eqnarray}
with the explicit form of the scattering matrices to be found in
appendix B.

The
boundary matrices will appear in an interesting  fashion in the TBA
as rapidity dependent fugacities.  For
example, for a state with $N$ pairs with rapidities
$\theta_1,...,\theta_N$, $N_q$ quartets with rapidities
$\theta^q_1,...,\theta^q_{N_q}$ and $N_b$ breather pairs with
rapidities $\theta^b_1,...,\theta^b_{N_b}$, the fugacity is
\begin{eqnarray}
\frac{\langle B_l | \alpha \rangle \langle \alpha | 
B_r \rangle}{\langle \alpha | \alpha \rangle} = \prod_{i=1}^N 
\bar{K}^{b_i a_i}_l (\theta_i) K^{a_i b_i}_r (\theta_i) 
\prod_{j=1}^{N_q} \det [\bar{K}_l (\theta^q_j) K_r(\theta^q_j)] 
\prod_{k=1}^{N_b} \bar{K}^{(n)}_l(\theta^b_k) K^{(n)}_r (\theta^b_k) 
\end{eqnarray}
In the thermodynamic limit, this expression can be rewritten in terms
of densities as
\begin{eqnarray}
\ln \frac{\langle B_l | \alpha \rangle \langle \alpha | 
B_r \rangle}{\langle \alpha | \alpha \rangle} &=& L \int_0^{\infty} d
\theta \left[
\sum_{ab} \ln [\bar{K}_l^{ba}(\theta) K_r^{ab}(\theta)] \rho_{ab}
(\theta) + \right. \nonumber \\
&&\left. + \ln \det [\bar{K}_l (\theta) K_r(\theta)] \rho_q (\theta) +
\sum_{n=1}^{\lambda-1} \ln [\bar{K}_l^{(n)} (\theta) K^{(n)}_r
(\theta)] \rho_n (\theta) \right].
\label{RLfugacity}
\end{eqnarray}
In a similar way, the energy can be written as
\begin{eqnarray}
E = L \int_0^{\infty} d \theta \left[ 2m \cosh \theta (\sum_{ab}
\rho_{ab} (\theta)  + 2 \rho_q (\theta)) + \sum_{n=1}^{\lambda-1} 2m_n
\cosh \theta ~\rho_n (\theta)\right],
\label{RLenergy}
\end{eqnarray}
whereas the entropy becomes in the Stirling approximation as
$L \rightarrow \infty$
\begin{eqnarray}
S \approx L \int_0^{\infty} d \theta \left[ (\sum_{ab} \rho_{ab} +
\rho_q + \rho_q + \rho^h) \ln (\sum_{ab} \rho_{ab} + \rho_q + \rho^h)
- \sum_{ab} 
\rho_{ab} \ln \rho_{ab} -
\right. \nonumber \\
\left.  
-\rho_q \ln \rho_q - \rho^h \ln \rho^h +
\sum_{n=1}^{\lambda -1} [(\rho_n + \rho_n^h) \ln (\rho_n + \rho_n^h) -
\rho_n \ln \rho_n - \rho_n^h \ln \rho_n^h ] \right].
\label{RLentropy}
\end{eqnarray}
Equations (\ref{RLfugacity}, \ref{RLenergy}, \ref{RLentropy}) can now
be substituted 
in the expression for the partition function in the R-channel,
(\ref{RchannelZagain}), which now becomes a functional integral over
the various densities.  The TBA equations then arise from the
saddle-point evaluation of this expression, subject to the constraints
coming from the quantization conditions
(\ref{RLquantizationconditions}).  Introducing the usual
quasienergies, we obtain after standard manipulations
\begin{eqnarray}
\epsilon_{ab} (\theta) &=& 2 m R \cosh \theta - \ln \bar{K}_l^{ba}
(\theta) K_r^{ab} (\theta) + \Phi * \ln \left[ 1 + \sum_{cd}
e^{-\epsilon_{cd} (\theta)} + e^{-\epsilon_q (\theta)} \right] + 
\sum_{n = 1}^{\lambda - 1} \Phi^{(n)} * \ln [ 1 + e^{-\epsilon_n
(\theta)}], \nonumber \\
\epsilon_q (\theta) &=& 4 m R \cosh \theta - \ln \mbox{det} \bar{K}_l
(\theta) K_r (\theta) + 2 \Phi * \ln \left[ 1 + \sum_{cd}
e^{-\epsilon_{cd} (\theta)} + e^{-\epsilon_q (\theta)} \right] + 
2 \sum_{n = 1}^{\lambda - 1} \Phi^{(n)} * \ln [ 1 + e^{-\epsilon_n
(\theta)}], \nonumber \\
\epsilon_n (\theta) &=& 2 m_n R \cosh \theta - \ln \bar{K}_l^{(n)}
(\theta) K_r^{(n)} (\theta) + \Phi^{(n)} * \ln \left[ 1 + \sum_{cd}
e^{-\epsilon_{cd} (\theta)} + e^{-\epsilon_q (\theta)} \right] + 
\sum_{m=1}^{\lambda -1} \Phi^{(n,m)} * \ln [ 1 + e^{-\epsilon_n
(\theta)}].
\end{eqnarray} 
Substituting these back in the expression for the partition function,
and 
absorbing the boundary factors into a redefinition of $\epsilon$'s, we arrive at the ground-state energy
\begin{eqnarray}
E_0 = -\frac{1}{2\pi} \int_0^{\infty} d \theta m\cosh \theta \ln
\left[1 + tr \bar{K}_l (\theta) K_r (\theta)
e^{-\epsilon(\theta)} + \det \bar{K}_l (\theta) K_r (\theta)
e^{-2\epsilon(\theta)} \right] - \nonumber \\
-\frac{1}{2\pi} \int_0^{\infty} d \theta \sum_{n=1}^{\lambda -1} m_n
\cosh \theta \ln \left[ 1 + \bar{K}^{(n)}_l (\theta) K^{(n)}_r (\theta)  
e^{-\epsilon_n(\theta)} \right]
\end{eqnarray}
where the energies are self-consistent solutions to the system of equations
\begin{eqnarray}
\epsilon (\theta) &=& 2m R \cosh \theta + \Phi * \ln \left[ 1 + tr
\bar{K}_l (\theta) K_r (\theta) e^{-\epsilon(\theta)} + 
\det \bar{K}_l (\theta) K_r (\theta) e^{-2\epsilon(\theta)} \right]  +
\nonumber \\ 
&&+ \sum_{n=1}^{\lambda -1} \Phi^{(n)} * \ln \left[ 1 + \bar{K}^{(n)}_l (\theta)
K^{(n)}_r (\theta) 
e^{-\epsilon_n(\theta)} \right] \nonumber \\
\epsilon_n (\theta) &=& 2 m_n R \cosh \theta + \Phi^{(n)} * \ln \left[1
+ tr \bar{K}_l (\theta) K_r (\theta)
e^{-\epsilon(\theta)} + \det \bar{K}_l (\theta) K_r (\theta)
e^{-2\epsilon(\theta)} \right] + \nonumber \\
&&+ \sum_{m=1}^{\lambda-1} \Phi^{(n,m)} * \ln \left[ 1 + \bar{K}^{(m)}_l
(\theta) K^{(m)}_r (\theta) e^{-\epsilon_m(\theta)} \right], 
\hspace{3cm} n = 1, ..., \lambda -1.
\end{eqnarray}

\subsection{Universal form}

These equations can be put in a simpler, universal form, following  
\cite{ZamolodchikovPLB253}.  The advantage of this reformulation is
that only one integral kernel remains.  For a given value of
$\lambda$, we label the breathers by $b = 1, ..., \lambda -1$, the
soliton by $b = \lambda$, and the anti-soliton by $b = \lambda + 1$.  We can
then use the remarkable identity \cite{ZamolodchikovPLB253}
\begin{eqnarray}
\left[ \delta_{ab} + \Phi_{ab} (k) \right]^{-1} = \delta_{ab} -
\frac{1}{2 \cosh k/h} L_{ab}
\end{eqnarray}
holding between the Fourier transform of the scattering kernels and
the incidence matrix $L_{ab}$ of the relevant Dynkin diagram (for
sine-Gordon at $\lambda$, this is the diagram of the $D_{\lambda + 1}$
algebra).  The Coxeter number is $h = 2 \lambda$ in that case.  

After basic manipulations, we obtain the universal form
\begin{eqnarray}
\epsilon_a (\theta) = \nu_a (\theta) + \sum_b L_{ab} \Phi_u * 
\left[ - \nu_b (\theta) + \epsilon_b (\theta) + \ln F_b (\theta)
\right], 
\end{eqnarray}
where the universal kernel is 
\begin{eqnarray}
\Phi_u (\theta) = \frac{\lambda}{2\pi \cosh \lambda \theta}
\end{eqnarray}
and the boundary dependence appears in the $F$ functions
\begin{eqnarray}
F_b (\theta) = \left \{ \begin{array}{cc} 
1 + \bar{K}_l^{(b)} (\theta) K_r^{(b)} (\theta) e^{-\epsilon_b
(\theta)}, & b = 1, ..., \lambda
- 1, \\
\left[ 1 + \mbox{tr} \bar{K}_l (\theta) K_r (\theta) e^{-\epsilon (\theta)}
+ \mbox{det} \bar{K}_l (\theta) K_r (\theta) e^{-2 \epsilon (\theta)}
\right]^{1/2}, & b = \lambda, \lambda + 1.
\end{array}
\right. 
\end{eqnarray}
in which we have defined the kink and antikink quasienergies as
$\epsilon_{\lambda} = \epsilon_{\lambda +1} = \epsilon$.
Finally, the driving terms $\nu$ are redundant
in the equations (the $\nu_a$ gets cancelled by the $\sum_b L_{ab}\Phi_u*[-\nu_b]$), but
very useful in the numerics, where they provide the correct
asymptotics at large rapidites.  They are
\begin{eqnarray}
\nu_a (\theta) = 2 m_a R \cosh \theta.
\end{eqnarray}

\subsection{The massless limit}

Taking the massless limit as described earlier leads to the following equations:
\begin{eqnarray}
E_0 = -\frac{1}{4\pi} \int_0^{\infty} d \kappa
\ln \left[1 + tr \bar{K}_l (\kappa) K_r (\kappa)
e^{-\epsilon(\kappa)} + \det \bar{K}_l (\kappa) K_r (\kappa)
e^{-2\epsilon f(\kappa)} \right] - \nonumber \\
-\frac{1}{4\pi} \sum_{n=1}^{\lambda -1} 
2 \sin{\frac{\pi n}{2\lambda}} \int_0^{\infty} d \kappa 
\ln \left[ 1 + \bar{K}^{(n)}_l (\kappa) K^{(n)}_r (\kappa)  
e^{-\epsilon_n(\kappa)} \right]
\end{eqnarray}
where the functions $\epsilon, \epsilon_n$  are self-consistent
solutions to the system of equations 
\begin{eqnarray}
\epsilon (\kappa)/\kappa &=& R + \int_0^{\infty} d \kappa' \Phi(\kappa,
\kappa') \ln \left[ 1 + tr
\bar{K}_l (\kappa') K_r (\kappa') e^{-\epsilon(\kappa')} + 
\det \bar{K}_l (\kappa') K_r (\kappa') e^{-2\epsilon (\kappa')} \right]  +
\nonumber \\ 
&&+ \sum_{n=1}^{\lambda -1} \int_0^{\infty} d \kappa' \Phi^{(n)}(\kappa,
\kappa') \ln \left[ 1 + \bar{K}^{(n)}_l (\kappa')
K^{(n)}_r (\kappa') 
e^{-\epsilon_n(\kappa')} \right] \nonumber \\
\epsilon_n (\kappa)/\kappa &=& 2 \sin{\frac{\pi n}{2\lambda}} R + \nonumber \\
&&+ \int_0^{\infty} d
\kappa' \Phi^{(n)} (\kappa, \kappa')  \ln \left[1
+ tr \bar{K}_l (\kappa') K_r (\kappa')
e^{-\epsilon(\kappa')} + \det \bar{K}_l (\kappa') K_r (\kappa')
e^{-2\epsilon (\kappa')} \right] + \nonumber \\
&&+ \sum_{m=1}^{\lambda-1} \int_0^{\infty} d \kappa' \Phi^{(n,m)}
(\kappa, \kappa') \ln \left[ 1 + \bar{K}^{(m)}_l
(\kappa') K^{(m)}_r (\kappa') e^{-\epsilon_m(\kappa')} \right], 
\hspace{0.5cm} n = 1, ..., \lambda -1.
\end{eqnarray}
The integral kernels are obtained from the massless limit of the known
ones, and are given explicitly by
\begin{eqnarray}
\Phi(\kappa, \kappa') &=& \frac{1}{\pi} \sum_{j=1}^{\lambda -1} \frac{\sin{\pi
j/\lambda}}{\kappa^2 + {\kappa'}^2 + 2 \kappa \kappa' \cos{\pi
j/\lambda}}, \nonumber \\
\Phi^{(n)} (\kappa, \kappa') &=& \frac{2}{\pi} \frac{(\kappa^2 +
{\kappa'}^2) \cos{\frac{\pi n}{2\lambda}}}{\kappa^4 + {\kappa'}^4 + 2
\kappa^2 {\kappa'}^2 \cos{\pi n/\lambda}} + \frac{2}{\pi}
\sum_{l=1}^{n-1} \frac{\cos{(n-2l)\frac{\pi}{2\lambda}}}{\kappa^2 +
{\kappa'}^2 - 2 \kappa \kappa' \sin{(n-2l)\frac{\pi}{2\lambda}}},
\nonumber \\
\Phi^{(n,m)} (\kappa, \kappa') &=& \frac{2}{\pi} (\kappa^2 +
{\kappa'}^2) \left[ \frac{\sin{(n+m)\frac{\pi}{2\lambda}}}{\kappa^4 +
{\kappa'}^4 - 2 \kappa^2 {\kappa'}^2 \cos{(n+m)\frac{\pi}{\lambda}}} + 
\frac{\sin{(n-m)\frac{\pi}{2\lambda}}}{\kappa^4 +
{\kappa'}^4 - 2 \kappa^2 {\kappa'}^2 \cos{(n-m)\frac{\pi}{\lambda}}}
\right] + \nonumber \\
&&+ \frac{2}{\pi} \sum_{l=1}^{m-1} \left
[ \frac{\sin{(m-n-2l)\frac{\pi}{2\lambda}}}{\kappa^2 + 
{\kappa'}^2 - 2 \kappa {\kappa'} \cos{(m-n-2l)\frac{\pi}{\lambda}}}
+ \frac{\sin{(m+n-2l)\frac{\pi}{2\lambda}}}{\kappa^2 + 
{\kappa'}^2 - 2 \kappa {\kappa'} \cos{(m+n-2l)\frac{\pi}{\lambda}}}
\right], \nonumber \\
&&\hspace{6cm} n \geq m.
\end{eqnarray}
The boundary scattering matrices become in the massless limit 
\begin{eqnarray}
K(\kappa) = \left[ \prod_{j=1}^{\lambda} \kappa +
\Delta^{\frac{\lambda + 1}{\lambda}}
e^{i\frac{\pi}{2\lambda} - i \frac{\pi j}{\lambda} + i \frac{\pi}{2}}
\right]^{-1} 
\left( \begin{array}{cc}
\kappa^{\lambda} & \Delta^{\lambda+1} e^{i\eta + i
\frac{\pi(\lambda-1)}{4}} \\
\Delta^{\lambda+1} e^{-i\eta + i \frac{\pi(\lambda-1)}{4}} &
\kappa^{\lambda} 
\end{array}
\right).
\end{eqnarray}
We have found it convenient to use a renormalizaed parameter $\Delta$ in this expression. What we  call $\Delta^{\lambda+1}$ coincides with what is called in other works \cite{FSW}  $T_B^\lambda$. This can be checked by comparing with the R matrix in the massless sine-Gordon model studied in \cite{FSW}, where the only difference with the present case is that the phase is missing. Now one knows
 from elsewhere the correspondence between the bare parameters $\tilde{\Delta}$ 
and $T_B$: 
\begin{equation}
T_B=\tilde{\Delta}^{\lambda+1\over\lambda}
{\Gamma(1/2\lambda)\over\Gamma(1/2)\Gamma(1/2+1/2\lambda)}\left[\sin{\pi\over\lambda+1}\Gamma(\lambda/(\lambda+1))\right]^{\lambda+1\over \lambda}
\end{equation}
We thus get the correspondence
\begin{equation}
\Delta=\tilde{\Delta}\left[{\Gamma(1/2\lambda)\over\Gamma(1/2)\Gamma(1/2+1/2\lambda)}\right]^{1/\lambda+1}
\left[\sin{\pi\over\lambda+1}\Gamma(\lambda/(\lambda+1))\right]^{1/\lambda}
\end{equation}
This reduces to $\Delta=\sqrt{\pi}\tilde{\Delta}$ in the case $\lambda=1, \beta^2=4\pi$ as already discussed. 

The additional piece of information that is not in \cite{FSW} is the relation between the phase in the $K$ matrix and the phase in the hamiltonian: one finds (see the apendix)
 $\eta=(\lambda+1)\chi$.

The trace and det of left and right $K$ matrices then read explicitly
(we have used $\eta_l = 0, \eta_r = (\lambda + 1) \chi$)
\begin{eqnarray}
tr \bar{K}_l (\kappa) K_r (\kappa) = 2 \left[ \prod_{j=1}^{\lambda} (\kappa +
\Delta_l^{\frac{\lambda + 1}{\lambda}}
e^{-i\frac{\pi}{2\lambda} + i \frac{\pi j}{\lambda} - i \frac{\pi}{2}})
(\kappa +
\Delta_r^{\frac{\lambda + 1}{\lambda}}
e^{i\frac{\pi}{2\lambda} - i \frac{\pi j}{\lambda} + i \frac{\pi}{2}})
\right]^{-1} \times \nonumber \\
\times \left[ \kappa^{2\lambda} + \Delta_l^{\lambda + 1}
\Delta_r^{\lambda+1} \cos{(\lambda+1) \chi} \right], \nonumber \\
det \bar{K}_l (\kappa) K_r (\kappa) =  \left[ \prod_{j=1}^{\lambda}
( \kappa +
\Delta_l^{\frac{\lambda + 1}{\lambda}}
e^{-i\frac{\pi}{2\lambda} + i \frac{\pi j}{\lambda} - i \frac{\pi}{2}}) 
(\kappa +
\Delta_r^{\frac{\lambda + 1}{\lambda}}
e^{i\frac{\pi}{2\lambda} - i \frac{\pi j}{\lambda} + i \frac{\pi}{2}})
\right]^{-2} \times \nonumber \\
\times \left[ \kappa^{2\lambda} + (-1)^{\lambda}
\Delta_l^{2(\lambda+1)}\right] \left[\kappa^{2\lambda} + (-1)^{\lambda}
\Delta_r^{2(\lambda+1)}\right].
\end{eqnarray}

The breather $K$ matrices are
\begin{eqnarray}
K^{(2n)}_{l,r} (\kappa) = - \prod_{l=1}^n \frac{\kappa^2 - 2 \kappa
\Delta_{l,r}^{\frac{\lambda +1}{\lambda}} \cos{\frac{(l-1/2)\pi}{\lambda}} +
\Delta_{l,r}^{2\frac{\lambda +1}{\lambda}}} {\kappa^2 + 2 \kappa
\Delta_{l,r}^{\frac{\lambda +1}{\lambda}} \cos{\frac{(l-1/2)\pi}{\lambda}} +
\Delta_{l,r}^{2\frac{\lambda +1}{\lambda}}} \nonumber \\
K^{(2n-1)}_{l,r} (\kappa) = \frac{\kappa - \Delta_{l,r}^{\frac{\lambda
+1}{\lambda}}}{\kappa + \Delta_{l,r}^{\frac{\lambda +1}{\lambda}}}
\prod_{l=1}^{n-1} \frac{\kappa^2 - 2 \kappa
\Delta_{l,r}^{\frac{\lambda +1}{\lambda}} \cos{\frac{l\pi}{\lambda}} +
\Delta_{l,r}^{2\frac{\lambda +1}{\lambda}}} {\kappa^2 + 2 \kappa
\Delta_{l,r}^{\frac{\lambda +1}{\lambda}} \cos{\frac{l\pi}{\lambda}} +
\Delta_{l,r}^{2\frac{\lambda +1}{\lambda}}}
\end{eqnarray}
Notice that the phase $ \chi$ appears only in the $K$ matrices for solitons and antisolitons, and for a single  boundary, could easily be gauged away 
by a redefinition of the soliton and antisoliton.

\section{The double boundary sine-Gordon model for  $\lambda = 2$}

\subsection{Generalities}

Let us now turn to a precise investigation of what the TBA equations entail.  
For the particular case $\lambda = 2$ (corresponding to $\beta^2 = 8\pi/3$, which is 
the next simplest case to 
tackle after the free fermion point $\lambda = 1$, $\beta^2 = 4\pi$), we can use the 
preceding results to rewrite the TBA in a 
more compact form.  We have explicitly (after rescaling to the more convenient 
dimensionless parameters $R
\Delta_{l,r}^{3/2} = \Delta_{L,R}^{3/2}$)
\begin{eqnarray}
E_0 &=& \frac{-1}{4\pi R} \int_0^{\infty} d \kappa \ln [1 + W_2 (\kappa)]
- \frac{\sqrt{2}}{4\pi R} \int_0^{\infty} d\kappa \ln \left[ 1 +
B_2 (\kappa) \right],
\nonumber \\
W_2 (\kappa) &=& \frac{\left[ 2(\kappa^4 + \Delta_L^3 \Delta_R^3
\cos{3\chi})e^{-\epsilon_2 (\kappa)} +  (\kappa^2 + \Delta_L^3 - \sqrt{2}
\kappa \Delta_L^{3/2})(\kappa^2 + \Delta_R^3 - \sqrt{2}
\kappa \Delta_R^{3/2}) e^{-2 \epsilon_2 (\kappa)} \right]}{(\kappa^2 + \Delta_L^3
+ \sqrt{2} \kappa \Delta_L^{3/2})(\kappa^2 + \Delta_R^3 + \sqrt{2}
\kappa \Delta_R^{3/2})}, \nonumber \\
B_2 (\kappa) &=& \frac{\kappa - \Delta_L^{3/2}}{\kappa + \Delta_L^{3/2}}
\frac{\kappa - 
\Delta_R^{3/2}}{\kappa + \Delta_R^{3/2}} e^{-\epsilon_1 (\kappa)}
\end{eqnarray}
with the quasienergies determined from the universal form 
of the massless TBA equations
\begin{eqnarray}
\epsilon_1 (\kappa) &=& \frac{2}{\pi} \int_0^{\infty} d \kappa' 
\frac{\kappa' \kappa^2}{{\kappa'}^4 + \kappa^4} \left\{ 2 \epsilon_2 (\kappa')
+ \ln [1 + W_2 (\kappa')] \right]\} \equiv
U ~\tilde{*} ~\left\{ 2 \epsilon_2 (\kappa)
+ \ln [1 + W_2 (\kappa)] \right\}, 
\nonumber \\
\epsilon_2 (\kappa) &=& \frac{2}{\pi} \int_0^{\infty} d \kappa' 
\frac{\kappa' \kappa^2}{{\kappa'}^4 + \kappa^4} \left\{ \epsilon_1 (\kappa')
+ \ln [1 + B_2 (\kappa')] \right] \equiv U ~\tilde{*} 
\left[ \epsilon_1 (\kappa)
+ \ln [1 + B_2 (\kappa)] \right\},
\end{eqnarray}
Here $U$ is the universal massless scattering kernel, and $\tilde{*}$
is the  ``massless convolution'' in the $\kappa$ variable.  

The solutions are required to obey the asymptotic conditions at $\kappa 
\rightarrow \infty$
\begin{eqnarray}
\lim_{\kappa \rightarrow \infty} \frac{\epsilon_1 (\kappa)}{\kappa} =
\sqrt{2}, \hspace{1cm}
\lim_{\kappa \rightarrow \infty} \frac{\epsilon_2 (\kappa)}{\kappa} =
1.
\end{eqnarray}
Numerically, the solution of the equations is straightforward in
the domain $0 \leq \chi < \pi/3$ (the sign of $\chi$ is not 
important).  In particular, all solutions $\epsilon_{1,2} (\kappa)$ 
are then real on the positive real axis of $\kappa$.   

Note that these TBA equations, 
from a knowledge of the functions on the real line, 
trivially define $\epsilon_{1,2} (\kappa)$ in the
domain $|\mbox{Arg} (\kappa)| \leq \pi/4$, which we call the first
fundamental domain.  This is due to the particular form of the 
integration kernel in the TBA equations, whose poles lie on the
aforementioned lines.  To obtain the TBA functions
outside of the first fundamental domain for a given value of $\chi$
from a knowledge of their values within the first fundamental domain,
one needs to perform an analytical continuation procedure (the details
of which are to be found below).  Note that as a function of $\chi$,
things will get even more complicated.  As we will see below, this will 
require another type of analytical continuation, with multiple steps,
generalizing the continuation proposed for the case $\lambda = 1$
in \cite{CauxPRL88}. 

The first bit of information we can gain comes from looking at the
limit $\kappa \rightarrow 0$ of the universal form.  Defining
$\lim_{\kappa \rightarrow 0} \epsilon_i = L_i$, we get the
equations
\begin{eqnarray}
L_1 = \frac{1}{2} \ln \left[ 1 + 2 \cos 3\chi e^{L_2} + e^{2 L_2}
\right], \hspace{1cm} L_2 = \frac{1}{2} \ln \left[ 1 + e^{L_1} \right].
\end{eqnarray}
Using the notation $x_i = e^{L_i}$, we can rewrite these as
\begin{eqnarray}
x_1 = x_2^2 - 1, \hspace{1cm} x_2 (x_2^2 - 3) = 2 \cos 3\chi.
\end{eqnarray}
These are useful equations allowing us to check the numerics.
However, these give us the first indication that not everything is
fine.  The third-order equation for $x_2$ yields three different 
solutions:  $x_2 = 2 \cos \chi, -\cos \chi \pm \sqrt{3}$.  
For $|\chi| \leq \pi/3$, the appropriate one to pick
is 
\begin{eqnarray}
x_2 = 2 \cos \chi.
\end{eqnarray}
At $\chi = \pi/3$, this therefore means that $x_1$ vanishes, and
therefore $L_1 \rightarrow -\infty$.  A naive solution to the TBA
equations would produce a cusp in $\chi$ here.  If one were to 
naively solve the TBA for $\chi$ beyond this value, one would
simply regenerate previous results:  for $\chi = \pi/3 + \eta$, 
the curves would be identical to those for $\chi = \pi/3 - \eta$,
which is patently incorrect in view of the cusp.  The same kind
of thing happened for $\lambda = 1$ at $\chi = \pi/2$,and we refer
the reader to the discussion in section 3 for a description
of the problems and their proper solution.

As we expect
analytic behaviour as a function of $\chi$, some form of analytic
continuation will have to be performed.  Note also that $x_2$ 
vanishes at $\chi = \pi/2$:  this further problem is not repaired
by the surgery at $\chi = \pi/3$, so the analytical continuation will
involve more than one step.  

To make progress on the problem, let us start by reformulating the
TBA equations into a more convenient form. Introduce the
following functions ($\omega = e^{i\pi/4}$)
\begin{eqnarray}
P(\kappa) &=& \frac{(\kappa - \Delta_L^{3/2})(\kappa - \Delta_R^{3/2})}
{(\kappa + \Delta_L^{3/2})(\kappa + \Delta_R^{3/2})}, \nonumber \\
Q(\kappa) &=& \left[\frac{(\kappa - \omega \Delta_L^{3/2})(\kappa - 
\bar{\omega} \Delta_L^{3/2})}{(\kappa + \omega \Delta_L^{3/2})(\kappa +
\bar{\omega} \Delta_L^{3/2})} \frac{(\kappa - \omega \Delta_R^{3/2})(\kappa - 
\bar{\omega} \Delta_R^{3/2})}{(\kappa + \omega \Delta_R^{3/2})(\kappa +
\bar{\omega} \Delta_R^{3/2})}\right]^{1/2},
\end{eqnarray}
and
\begin{eqnarray}
\cos \theta(\kappa,\chi) = \frac{\kappa^4 + \Delta_L^3 \Delta_R^3 
\cos 3\chi}{(\kappa^4 + \Delta_L^6)^{1/2} (\kappa^4 + \Delta_R^6)^{1/2}}.
\end{eqnarray}

The eigenvalues of $\bar{K}_{l}K_{r}$ are equal to $Q(\kappa)
e^{\mp i \theta(\kappa,\chi)}$. Introducing then the $Y$ functions
through
\begin{eqnarray}
e^{\epsilon_1}(\kappa) \equiv P (\kappa) Y_1 (\kappa), \hspace{1cm} 
e^{\epsilon_2} (\kappa) \equiv Q (\kappa) e^{\mp i \theta(\kappa,\chi)} 
Y_2^{\pm} (\kappa)
\end{eqnarray}
the universal form of the TBA can be rewritten
\begin{eqnarray}
\ln Y_1 (\kappa) &=& - \ln P(\kappa) + U ~\tilde{*} 
~\left\{ \ln Q^2 (\kappa)
+ \ln [1 + Y_2^+ (\kappa')] + \ln [1+Y_2^- (\kappa')] \right\},
\nonumber \\
\ln Y_2^{\pm} (\kappa) &=& \pm i \theta(\kappa,\chi) - \ln Q (\kappa) +
U ~\tilde{*} ~\left\{
\ln P(\kappa) + \ln [1 + Y_1 (\kappa)] \right\}.
\end{eqnarray}
We can in fact verify 
the following integrals, for $d, \kappa \in \mathbb{R}^+$:
\begin{eqnarray}
\frac{2}{\pi} \int_0^{\infty} d \kappa' \frac{\kappa' \kappa^2}
{{\kappa'}^4 + \kappa^4} \ln \left( \frac{\kappa' - \omega d}
{\kappa' + \omega d} \frac{\kappa' - \bar{\omega} d}{\kappa' 
+ \bar{\omega} d} \right) &=& \ln \frac{\kappa^2 + d^2}{(\kappa
+ d)^2}, \nonumber \\
\frac{2}{\pi} \int_0^{\infty} d \kappa' \frac{\kappa' \kappa^2}
{{\kappa'}^4 + \kappa^4} \ln \left( \frac{\kappa' - d}
{\kappa' + d} \right) &=& \frac{1}{2} \ln \left( \frac{\kappa - \omega d}
{\kappa + \omega d} \frac{\kappa - \bar{\omega} d}{\kappa 
+ \bar{\omega} d} \right) \pm i \arctan \frac{d^2}{\kappa^2}
\end{eqnarray}
where we have to make a branch choice in the last equation.  In fact,
knowing that the phase of $Y_2^{\pm}$ is given by $\pm \theta$, 
we can see that the appropriate choice should yield a 
simplification of the
TBA equations with the identities
\begin{eqnarray}
U ~\tilde{*} ~\ln Q^2 (\kappa)&=&
\ln P(\kappa) + \ln R (\kappa), \nonumber \\
U ~\tilde{*} ~\ln P (\kappa) 
&=& \ln Q (\kappa),
\end{eqnarray}
where 
\begin{eqnarray}
R (\kappa) = \frac{\kappa^2 + \Delta_L^3}{\kappa^2 - \Delta_L^3}
\frac{\kappa^2 + \Delta_R^3}{\kappa^2 - \Delta_R^3}.
\end{eqnarray}
The final form of the TBA is therefore
\begin{eqnarray}
\ln Y_1 (\kappa) &=& \ln R (\kappa) + U ~\tilde{*} 
~\left\{ \ln [1 + Y_2^+ (\kappa)] + 
\ln [1+Y_2^- (\kappa)] \right\},
\nonumber \\
\ln Y_2^{\pm} (\kappa) &=& \pm i \theta(\kappa, \chi)
 + U ~\tilde{*} ~\ln [1 + Y_1 (\kappa)].
\end{eqnarray}
The ground state energy then reads
\begin{eqnarray}
E_0 = \frac{-1}{4\pi R} \int_0^{\infty} d \kappa \left\{
\sqrt{2} \ln [1 + Y_1^{-1} (\kappa)] + \ln [1 + {Y_2^{+}}^{-1} (\kappa)]
+ \ln [1 + {Y_2^-}^{-1} (\kappa)] \right\}.
\label{GSEYsystem}
\end{eqnarray}

Note that, one has to be careful along the lines
$\mbox{arg}(\kappa) = \pm \pi/4$ where the universal kernel leads to 
singular integrals, which have to be evaluated by contour deformation.
First, note that simple manipulations yield
\begin{eqnarray}
P (\omega \kappa) P (\bar{\omega} \kappa) = Q^2 (\kappa), 
\hspace{1cm} Q (\omega \kappa) Q (\bar{\omega} \kappa) =
P (\kappa).
\label{PQperiodicity}
\end{eqnarray}
and also $R(\omega \kappa) = 
R^{-1} (\bar{\omega} \kappa)$.
Observing that $\theta(\omega \kappa, \chi) = \theta(\bar{\omega} 
\kappa, \chi)$ for real $\kappa$, 
and considering the evaluation of the TBA functions on these lines, 
yields the $Y$-system
\begin{eqnarray}
Y_1 (\omega \kappa) Y_1 (\bar{\omega} \kappa) &=& [1 +
Y_2^+ (\kappa)][1 + Y_2^- (\kappa)], \nonumber \\
Y_2^{\pm} (\omega \kappa) Y_2^{\mp} (\bar{\omega} \kappa) &=& 
1 + Y_1 (\kappa)
\end{eqnarray}
which coincides with the $D_3$ $Y$-system in \cite{ZamolodchikovPLB253}.  In particular, the nontrivial
periodicity in the complex plane of $\kappa$ can be read from those.
Namely, one can show by direct substitution that
\begin{eqnarray}
Y_1 (\omega^6 \kappa) = Y_1 (\kappa), \hspace{1cm} 
Y_2^{\pm} (\omega^6 \kappa) = Y_2^{\pm} (\kappa).
\end{eqnarray}
This means that the $Y$-functions naively 
have Laurent expansions in powers of
$\kappa^{4/3}$.  In particular, the boundary pairings appear in
this expansion in powers of $\Delta_{L,R}^2$.  Once again, like at
the free fermion point, the TBA does not see clearly the term 
proportional to $\Delta_L \Delta_R$ in the perturbative expansion
of the ground-state energy at weak boundary pairing.

In view of the periodicity of the $Y$ functions, and the consequence
this has on the structure of their Laurent expansions, we see that the
ground-state energy has a naive expansion in powers of $\Delta_{L,R}^2$.
However, as we have stated above, problems are encountered at
$\chi = \pi/3$ which require modifying the TBA in a nontrivial way.
We now discuss how this surgery is to be performed to ensure
analyticity in the boundary phase difference $\chi$.

\subsection{Scenario for the analytic continuation}
As we have argued above, 
we encounter various types of singularities in the 
(solutions to the) TBA equations as a function of $\chi$.  
Let us define the
following terminology:  types $K_1, K_2^{\pm}$ 
singularities are defined for $\kappa$ obeying
\begin{eqnarray}
1 + Y_1 (\kappa) = 0, \hspace{1cm} 1+Y_2^{\pm} (\kappa) = 0.
\end{eqnarray}
We also define types $F_1, F_2^{\pm}$ by
\begin{eqnarray}
Y_1 (\kappa) = 0, \hspace{1cm} Y_2^{\pm} (\kappa) = 0,
\end{eqnarray}
and types $G_1, G_2^{\pm}$ by
\begin{eqnarray}
Y_1^{-1} (\kappa) = 0, \hspace{1cm} (Y_2^{\pm})^{-1} = 0.
\end{eqnarray}
Types $K_1, K_2^{\pm}$ and $F_1, F_2^{\pm}$ have consequences
on the ground-state energy integral, whereas $G_1, G_2^{\pm}$
types are harmless.  

We now have to ask ourselves what singularities appear in
our equations, and how we have to take them into account.  
From the numerical solution of the TBA equations, we know
that at $\chi = \pi/3$, there are singularities of types
$K_2^{\pm}$ and $F_1$ at $\kappa = 0$.  We propose the
following scenario for the analytical continuation of the
TBA beyond $\chi = \pi/3$.

First of all, notice that a singularity of type $K_2^{\pm}$
(as we know appears at $\chi = \pi/3$) requires $\ln Y_2^{\pm}
 = i \pi ~\mbox{mod}(2\pi)$.  In particular, the real part of
$\ln Y_2^{\pm}$ must vanish (this is equivalent to saying that
$\epsilon_2$ must be purely imaginary, as was argued for
the case $\lambda = 1$ at the end of section 3).
By examining the form of the
universal scattering kernel, and in view of the comments made
earlier about the use of the TBA equations to write the functions
$Y$ in the complex plane of $\kappa$ from a knowledge of their
value on the real line, we can easily convince ourselves that
the condition $\Re{~\ln Y_2^{\pm}} = 0$ could only be satisfied
on the axes $\omega \kappa, \bar{\omega} \kappa$ with $\kappa
\in \mathbb{R}$.  We therefore state

Proposition 1:  at $\chi = \pi/3$, 
singularities of type $K_2^{\pm}$ and $F_1$ occur at $\kappa = 0$.
Increasing $\chi$ beyond $\pi/3$ moves the $K_2^{\pm}$ singularities
to $\omega \kappa_s$, $\bar{\omega} \kappa_s$, with $\kappa_s 
\in {\mathbb{R}} > 0$.  The $F_1$ singularity moves to $\kappa_s$
(that is, the $K_2^{\pm}$ singularities are at angles $\pm \pi/4$
in the complex plane of $\kappa$, and the $F_1$ singularity sits
on the real line). 
Note that the relative position of the singularities is dictated
by the $Y$-system.  

If such a singularity structure occurs, 
the first TBA equation
needs to be modified to take the pole contribution into account:
\begin{eqnarray}
\ln Y_1 (\kappa) = \ln T(\kappa, \kappa_{a1}) + 
\ln R_1 (\kappa) 
+ U ~\tilde{*} ~\left\{ \ln [1 + Y_2^+ (\kappa)] + 
\ln [1+Y_2^- (\kappa)] \right\},
\label{univcorrected}
\end{eqnarray}
where we have defined the function
\begin{eqnarray}
\ln T (\kappa, \kappa_s) = \ln \frac{\kappa^2 - \kappa_s^2}
{\kappa^2 + \kappa_s^2} 
\end{eqnarray}
as the residue that pops out of the universal kernel.  Note that 
this modification ensures the presence of a $F_1$ singularity
on the real line at $\kappa_s$, as required.  Note also that the
extra term does not change the $Y$-system structure, since 
$T(\omega \kappa, \kappa_s) T(\bar{\omega} \kappa, \kappa_s) = 1$.

The ground state energy changes under this continuation:  we will
treat this after solving the singularity problem of the TBA 
equations.

The major remaining problem is to determine the value of
$\kappa_s$.  This comes from solving the singularity condition 
obtained by explicitly requiring e.g. $Y_2^{+} (\omega \kappa_s) = -1$
under Proposition 1.  This translates into
\begin{eqnarray}
\pi ~(\mbox{mod} ~2\pi) = \theta(\omega \kappa_s, \chi) + 
\frac{2}{\pi} P \int_0^{\infty} d\kappa' \frac{\kappa' 
\kappa_s^2}{{\kappa'}^4 - \kappa_s^4} 
\ln [1 + Y_1 (\kappa')].
\end{eqnarray}
Note that the fact that $Y_1$ has a type $F_1$ singularity at
$\kappa_s$ makes the principal part integral well-behaved.
The TBA system now comprises three equations, i.e. the two basic
TBA equations, with the addition of the first singularity condition.

The first thing to notice is that this modification cures 
our problems at $\chi = \pi/3$ with the $\kappa \rightarrow 0$ asymptotics.
That is, the relevant formulas become
\begin{eqnarray}
L_1 = \ln (-1) + 
\frac{1}{2} \ln \left( 1 + 2 \cos 3\chi e^{L_2} + e^{2 L_2}
\right), \hspace{1cm} L_2 = \frac{1}{2} \ln \left( 1 + e^{L_1} \right) 
\end{eqnarray}
Again using the notation $x_i = e^{L_i}$, we can rewrite these as
\begin{eqnarray}
x_1 = 1 - x_2^2, \hspace{1cm} x_2 (x_2^2 - 3) = 2 \cos 3\chi.
\end{eqnarray}
providing a smooth curve for $x_1$ (note the change of sign)
around the singular point.

Now as we tune $\chi$ up beyond this, we encounter the possibility
of a type $K_1$ singularity, so this is by far not the end of 
the story.  From the asymptotics at $\kappa 
\rightarrow 0$, we expect this to happen at
$\chi = \pi/2$, where $x_2$ is bound to change sign (meaning that
the function $Y_2$ must become negative on at least part of the
real axis of $\kappa$).  
We therefore 
expect a $K_1$ singularity at $\kappa = 0$ for $\chi = \pi/2$.

Moreover, a careful study of the first 
singularity
condition above shows that one can solve it for $\kappa_s$ real
only provided (here, for $\Delta_L = \Delta_R$)
\begin{eqnarray}
\kappa_s \leq \Delta^{3/2} \sqrt{|\cos \frac{3\chi}{2}|}.
\end{eqnarray}
The immediate consequence of this is that the window of 
possible values of the first singular point grows between $\chi = \pi/3$
and $\pi/2$, but decreases again afterwards.  We therefore expect the
first singularity to collapse back to zero before $\chi$ reaches $\pi$.

If such a second singularity crosses the real
axis, then we have to modify the TBA again.  Namely, we now get
a modified second TBA equation:
\begin{eqnarray}
\ln Y_2^{\pm} (\kappa) = \pm i \theta(\kappa, \chi) + 
\ln T(\kappa, \kappa_{s2}) + 
U ~\tilde{*} ~\ln [1 + Y_1 (\kappa)]
\end{eqnarray}
where the parameter $\kappa_{s2}$ is a solution to 
the second type of singularity condition
\begin{eqnarray}
\pi = i \ln T(\omega \kappa_{s2}, \kappa_{s}) -i \ln R_1 
(\omega \kappa_{s2})
- \frac{2}{\pi} P \int_0^{\infty} d \kappa' 
\frac{\kappa' \kappa_{s2}^2}{{\kappa'}^4 - \kappa_{s2}^4}
\left[ \ln [1 + Y_2^+ (\kappa')] + 
\ln [1 + Y_2^- (\kappa')] \right]
\end{eqnarray}

\begin{center}
\begin{figure}
\parbox{10cm}{
\centerline{\epsfig{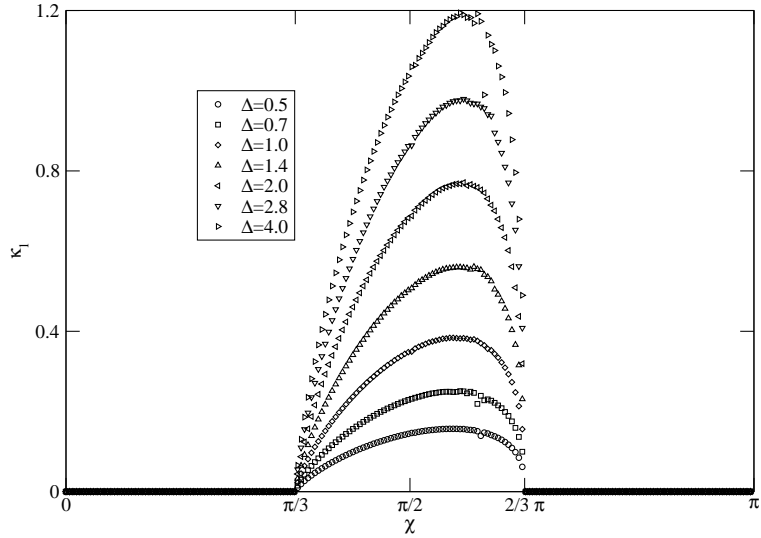}}}
\vspace{1.0cm}
\caption{The singularity $\kappa_{s1}$ 
plotted as a function of the phase difference $\chi$ for seven values of
the boundary parameters $\Delta_L = \Delta_R$.}
\label{k1.allD}
\end{figure}
\end{center}

\begin{center}
\begin{figure}
\parbox{8cm}{
\centerline{\epsfig{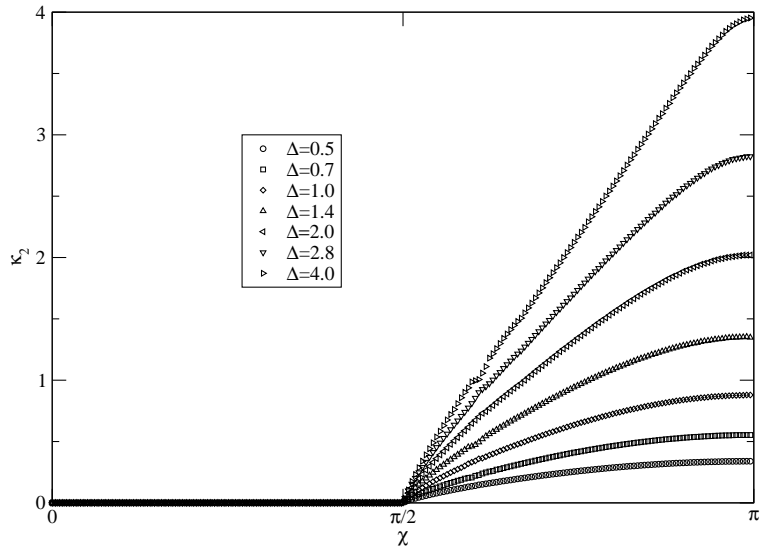}}}
\vspace{1.0cm}
\caption{The singularity $\kappa_{s2}$ 
plotted as a function of the phase difference $\chi$ for seven values of
the boundary parameters $\Delta_L = \Delta_R$.}
\label{k2.allD}
\end{figure}
\end{center}

The final modification from the second singularity occurs in
the first singularity condition:  namely, the integral kernel
appearing in it again diverges, and has to be continued.  
This produces the same king of additional contribution as
appeared in the TBA equations, corresponding to the residue
of the universal scattering kernel evaluated at appropriate
points.  For
clarity, we rewrite the full TBA, including the appropriate 
branch choices, for the case $\lambda = 2$.  In this, we have
found it convenient to reabsorb the phases $\pm i \theta$ and
the kernels $T$ in the $Y$-functions.  The full TBA finally reads
\begin{eqnarray}
\ln Y_1 (\kappa) &=& \ln R(\kappa) + U \tilde{*} 
\ln \left[1 + 2 \cos \theta (\kappa) \frac{\kappa^2 - \kappa_{s2}^2}
{\kappa^2 + \kappa_{s2}^2} Y_2 (\kappa) + \left(\frac{\kappa^2 - \kappa_{s2}^2}
{\kappa^2 + \kappa_{s2}^2}\right)^2 Y_2^2 (\kappa) \right], \nonumber \\
\ln Y_2 (\kappa) &=& U \tilde{*} \ln \left[ 1 +  \frac{\kappa^2 - \kappa_{s1}^2}
{\kappa^2 + \kappa_{s1}^2} Y_1 (\kappa) \right], 
\end{eqnarray}
supplemented by the two singularity conditions
\begin{eqnarray}
\pi &=& \arccos{\frac{\Delta_L^3 \Delta_R^3 \cos 3\chi - \kappa_{s1}^4}
{\sqrt{(\Delta_L^3 - \kappa_{s1}^4)(\Delta_R^3 - \kappa_{s1}^4)}}}
+ 2 \arccos{\frac{\kappa_{s1}^2}{\sqrt{\kappa_{s1}^4 + \kappa_{s2}^4}}} 
+ \frac{2}{\pi} \mbox{P}\int_0^{\infty} d \kappa' 
\frac{\kappa' \kappa_{s1}^2}{{\kappa'}^4 - \kappa_{s1}^4} \ln \left[
1 + \frac{{\kappa'}^2 - \kappa_{s1}^2}{{\kappa'}^2 + \kappa_{s1}^2}
Y_1 (\kappa') \right] , \nonumber \\
\pi &=& -2 \arccos{\frac{\kappa_{s2}^2}{\sqrt{\kappa_{s1}^4 + \kappa_{s2}^4}}}
+ 2 \arccos{\frac{\kappa_{s2}^2}{\sqrt{\kappa_{s2}^4 + \Delta_L^6}}}
+ 2 \arccos{\frac{\kappa_{s2}^2}{\sqrt{\kappa_{s2}^4 + \Delta_R^6}}}
- \nonumber \\
&&- \frac{2}{\pi} \mbox{P}\int_0^{\infty} d \kappa' 
\frac{\kappa' \kappa_{s1}^2}{{\kappa'}^4 - \kappa_{s1}^4} \ln \left[
1 + 2 \cos \theta (\kappa') \frac{{\kappa'}^2 - \kappa_{s2}^2}
{{\kappa'}^2 + \kappa_{s2}^2} Y_2 (\kappa') + \left(\frac{{\kappa'}^2 - \kappa_{s2}^2}
{{\kappa'}^2 + \kappa_{s2}^2}\right)^2 Y_2^2 (\kappa') \right],
\end{eqnarray}
with $\kappa_{s1} = 0$ for $\chi < \pi/3$ and $\chi > 2\pi/3$, and
$\kappa_{s2} = 0$ for $\chi < \pi/2$.  The evolution of the singularities
is as follows:  $\kappa_{s1}$ first appears at $\chi = \pi/3$, and moves
up.  At $\chi = \pi/2$, $\kappa_{s2}$ makes its appearance, and starts
moving up.  As it does so, it pulls $\kappa_{s1}$ back down, until the
latter vanishes again at $\chi = 2\pi/3$.  After that point, only
$\kappa_{s2}$ remains.  This behaviour is illustrated in figures
(\ref{k1.allD}, \ref{k2.allD}), where the two singularities are plotted
as a function of the phase difference $\chi$ for various values of 
the boundary parameters $\Delta_L = \Delta_R$.

\vspace{1cm}
\vspace{1cm}

The ground state energy is given in these notations 
by the integral expression (\ref{GSEYsystem}) analytically continued
for the singularities of type $K_2^{\pm}$ and $F_1$ at 
$\chi = \pi/3$, and those of type $K_1$ and $F_2^{\pm}$ at
$\chi = \pi/2$.  The extra kernels from these respective
divergences cancel pairwise, and one is left after a few basic
manipulations with the modified integral expression for the
ground state energy (valid for all values of $\chi$)
\begin{eqnarray}
E_0 &=& \frac{-1}{4\pi R} \int_0^{\infty} d \kappa \left\{
\ln \left[ 1 + 2 \cos \theta (\kappa) \frac{{\kappa}^2 - \kappa_{s2}^2}
{{\kappa}^2 + \kappa_{s2}^2} Y_2 (\kappa) 
+ \left(\frac{{\kappa}^2 - \kappa_{s2}^2}
{{\kappa}^2 + \kappa_{s2}^2}\right)^2 Y_2^{2} (\kappa) \right]
- 2 \ln Y_2(\kappa) + \right. \nonumber \\
&& + \left. \sqrt{2} 
\ln \left[1 + \frac{{\kappa}^2 - \kappa_{s1}^2}{{\kappa}^2 + \kappa_{s1}^2}
Y_1 (\kappa) \right] - \sqrt{2} \ln Y_1(\kappa)\right\}.
\end{eqnarray}
Due to the fact that we have reabsorbed the singularity contributions into
the $Y$ functions, they are always real and positive on the real line.  

The numerical solution to the full TBA equations (including the singularity
conditions) over the full interval of $\chi$ from $0$ to $\pi$ turns out
to be a considerable challenge.  The evaluation of the principal part 
integrals requires many sampling points, but this isn't the main problem.
The main problem is that the numerics tend to be unstable within a small 
domain around $\chi \approx 1.9$.  This is due to the fact that the
right-hand side of the singularity conditions are nonmonotonous functions,
from which it is therefore difficult to isolate the correct root.  Four
integral equations have to be simultaneously obeyed, so the iteration
procedure in general has many unstable directions.  
A simple root finding algorithm is insufficient, and one has to make the
code rather elaborate to obtain sensible curves.

\begin{center}
\begin{figure}
\parbox{8cm}{
\centerline{\epsfig{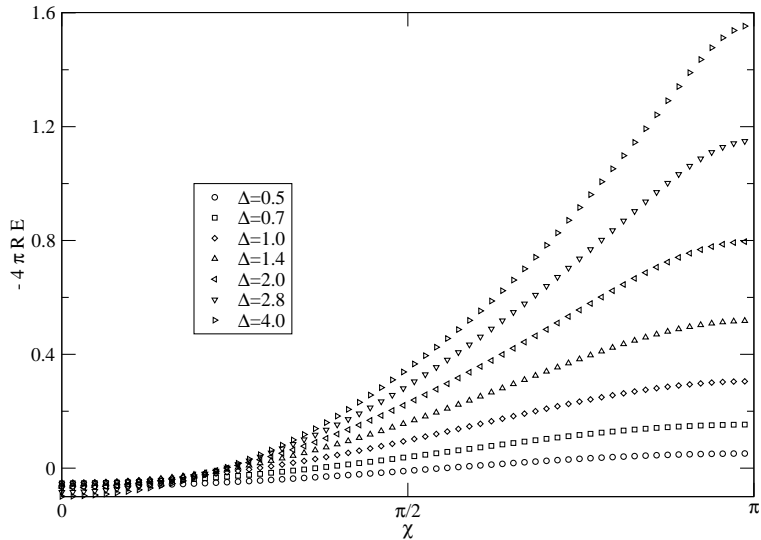}}}
\vspace{1cm}
\caption{Ground-state energy of the two-boundary sine-Gordon model
plotted as a function of the phase difference $\chi$ for seven values of
the boundary parameters $\Delta_L = \Delta_R$.  The interpolation between
the conformal limits is clearly seen.}
\label{GSE.allD}
\end{figure}
\end{center}

The curves for the ground-state energy
as a function of $\chi$ for various values of the boundary parameters 
$\Delta$ are presented in figure (\ref{GSE.allD}).  
>From these, it is transparent that
our construction provides a completely interpolating solution between the
$\sim \cos \chi$ behaviour at small $\Delta$ to the $\sim \chi^2$ one for
large $\Delta$, as expected from the appropriate conformal limits.
For large $\Delta$, the fit is perfect with the expected functional form
(easily derived from mode expansions)
\begin{eqnarray}
E_0 (\Delta \rightarrow \infty) = -\frac{\pi}{24R} 
+ \frac{2 \chi^2}{\beta^2 R} = \frac{\pi}{24R} + \frac{3\chi^2}{4\pi}
\end{eqnarray}
with the last equality valid for the present case $\lambda = 2$.

\vspace{1cm}

\subsection{Some remarks on the general singularity structure}

We are not sure what the general singularity structure will be for other 
values of $\lambda$. A generalization of the Ising model argument given 
in section II indicates that there always will be a singularity at ${\pi\over 2}$.
Indeed, following \cite{GhoshalIJMPA9},
boundary states are properly understood by 
considering a semi-infinite system, and choosing at infinity either the boundary 
condition $\phi=0$ or $\phi={2\pi\over\beta}$ (the equivalent of spin up or down 
in the Ising model). For $\chi\in\left[0,{\pi\over 2}\right]$, the first condition 
defines the true ground state (equivalent to $h>0$ in the Ising model) while for 
$\chi\in\left[{\pi\over 2},\pi\right]$, the first condition corresponds to an 
excited state. Equivalently, if one sets always $\phi=0$ at infinity, then the 
case $\chi\in\left[{\pi\over 2},\pi\right]$ should correspond to an excited boundary state.  

We pause here to stress that the ground state energy does not have any singularity, 
and that there is no level crossing the ground state at any of the singularity values 
in the massless,  two boundary problem. Indeed, the foregoing argument assumed values 
at infinity fixed by the bulk interaction term, with periodicity 
$\phi\rightarrow \phi+{2\pi\over\beta}$. In the massless two boundary problem, the 
periodicity is $\phi\rightarrow\phi+{4\pi\over\beta}$, so the first level crossing 
is expected to occur at $\chi=\pi$. 

The existence of the singularities we are struggling with translates into interesting 
behaviours for the excited levels of the hamiltonian. This can be seen quite clearly 
in the case of the Ising model, for instance 
 with boundary fields $h_l=h$ fixed, $h_r=h'$ varying. 
Since the ground state is analytic, the first gap  behaves as ${1\over 2}|hh'|$ at 
small magnetic fields, and pinches the ground state in a V shape around the symmetric point $h'=0$.

Technically, the existence of singularities in the TBA approach follows from the 
dependence of the scattering matrices on $(\lambda+1)\chi$ instead of $\chi$. This 
dependence is a consequence of the fact that the scattering matrices are in essence IR defined. 
Starting from the UV action (\ref{basic1}) the Dirichlet IR boundary conditions 
are approached along the operators ``dual'' to $e^{i\beta\phi/2}$,  i.e. 
$e^{i\beta_d\phi_d/2}$ where $\phi_d=\phi-\bar{\phi}$
and ${\beta_d^2\over 8\pi}={8\pi\over \beta^2}$. It follows that the dual coupling 
$\tilde{\Delta}_d\propto \left(\tilde{\Delta}\right)^{-\lambda-1}$ and therefore, 
if $\chi$ is the phase of the UV coupling, the phase of the IR coupling becomes 
$(\lambda+1)\chi$.  Clearly, the IR action (or the scattering matrix)  by itself 
is not sufficient to describe all values of $\chi\hbox{ Mod }2\pi$, and it is natural 
to expect that the reduced information contained in $(\lambda+1)\chi\hbox{ Mod }2\pi$ 
has to be supplemented with specification of a boundary state. While it is possible 
to associate a conformal Dirichlet boundary state with any value of the field 
$\phi(0)$, it is well known (see eg 
\cite{AOS}, whose conventions we follow)
that for a given radius $R={2\over\beta}$, there are, when $R^2={\lambda+1\over 2\pi}$, 
$\lambda+1$ ``special'' boundary states. Presumably, the IR action (or the scattering 
matrix) supplemented by the choice of one of these boundary states would allow to 
explore the whole domain $\chi\hbox{ Mod }2\pi$, which would correspond  in the TBA 
approach to  analytic continuation at the values $\chi={n\pi\over\lambda+1}$. 

It may also be that more values require continuation. This can be inferred from the 
limit $\kappa\rightarrow 0$, which gives the following values of the $x_j$ for general 
$\lambda$: $x_j={\sin(\lambda+2-j)\chi\over\sin\chi}$. We thus expect from this  to 
observe singularities also at values ${\pi\over\lambda+1},{\pi\over\lambda},{\pi\over 2}$, 
and multiples (the pattern is similar but different from the one of excited boundary states 
discussed in \cite{Takacs}).More work is required to clarify this question.

\section{The double Kondo model}

\bigskip

The TBA approach to the double Kondo model is bound to be rather intricate, as the boundary scattering involves now (except for spin $1/2$) boundary degrees of freedom, whose inclusion into the boundary state and the subsequent R-channel TBA is not obvious. Fortunately, the problem is perfectly well suited, on the other hand, to the Destri de Vega approach. 

The general framework to apply the DDV approach to a problem with two boundaries has been discussed in details in \cite{LeClairNPB453}. There, the starting point was the inhomogeneous 6 vertex model with boundary fields. The role of the inhomogeneities was to introduce a bulk mass scale, while the boundary fields constrained the value of the sine-Gordon field on the boundaries. The net result was an expression for the ground state energy 
of the  sine-Gordon model with double Dirichlet boundary conditions.

It is well known how to introduce a different kind of inhomogeneities in the 6 vertex model to get a theory that is massless, with free L movers and R movers interacting with a Kondo type impurity. It is similarly possible to obtain a theory with free L movers and R movers interacting with two different Kondo type impurities. Forgetting about the L movers, one can then transform this problem by folding into the problem we are interested in. It is a straightforward calculation to then obtain expressions for the ground state energy - these were actually almost already written in \cite{LeClairNPB453}\footnote{There is a misprint in this paper, as equation (9.3) corresponds to Neumann, not Dirichlet boundary conditions on the other boundary.}. Since the most interesting limit is the isotropic  one (XXX model), we give results in the repulsive regime of the sine-Gordon model, parametrizing $\beta^2=8\pi {t-1\over t}$. 
One finds then
\begin{eqnarray}
f(\theta) &=& i\mu e^\theta+iP_{bdry}(\theta)
-2i\int_{-\infty}^\infty d\theta'\Phi(\theta-\theta')\hbox{Im }\ln\left[1-e^{f(\theta'+i0)}\right]\nonumber\\
E &=& -{2 \over \pi R}\int_{-\infty}^\infty \mu Re^\theta
\hbox{Im }\ln\left[1-e^{f(\theta+i0)}\right]
\end{eqnarray}
where $\Phi$ is the kernel already encountered in section IV
\begin{equation}
\Phi(\theta)=-\int_{-\infty}^\infty {dx\over 2\pi^2}
{\sinh(t-2)x\over\cosh x\sinh(t-1)x}e^{2ix\theta/\pi}
\end{equation}
and the boundary kernel is
\begin{equation}
iP_{bdry}(\theta)=i\int_{-\infty}^\infty {dx\over x} {\sinh (t-j_l)x\over 2\cosh x\sinh(t-1)x} \sin 2x(\theta-\theta_{Bl})/\pi -{i\pi\over 2}{t-j_l\over t-1} +(l\rightarrow r)
\end{equation}
The phase is such that at small coupling $\theta_B\rightarrow -\infty$ the boundary term vanishes while at large coupling  $\theta\rightarrow\infty$ it reads $iP_{bdry}=\pi\left({t-j_l\over t-1}+{t-j_r\over t-1}\right)$, this maybe modulo $2\pi$. 

Note the deep similarity of these equations with the ones usually written in the bulk - in the latter case, there would be a $i\omega$ term instead of a $iP_{bdry}$ term, the $\omega$ corresponding to a soliton fugacity. 

While bulk equations involve a constant $\omega$, the variation of $P_{bdry}$ with the rapidity leads to some obvious difficulties. Let us consider for instance the simplest case of a spin $1/2$ Kondo impurity at one boundary, Neumann conditions at the other boundary, and restrict also to the isotropic case $t=\infty$. As discussed in  \cite{Affleck} going from UV to IR in this case amounts to fusing the identity
(Kac Moody) representation that initially goes through the system with the spin $1/2$ (Kac Moody) representation, so the ground state energy in the IR should correspond to the conformal weight $h={1/2(1/2+1)/1+2}=1/4$. This is in agreement with our formula, as the usual dilogarithm analysis gives
\begin{equation}
h_{IR}={1\over 4}\left({P_{bdry}(\infty)\over\pi}\right)^2={1\over 4}
\end{equation}
If however we put two Kondo impurities, we see that our formula gives $h_{IR}=1$, while fusing with the Kac Moody twice gives the Kac Moody identity representation back, with lowest conformal weight $h=0$. The $h=1$ weight is instead an excited weight, and therefore our DDV equation, while it captures the ground state at small coupling, captures instead an excited state at large coupling. 

It was also argued by Affleck and Ludwig in \cite{AffLu} that the picture for $j={1\over 2}$ essentially carries over to the case of arbitrary $j$: in either case, flow from UV to IR is described by the absorption of an eletron by the impurity, and a phase shift of $\pi/2$ for the fermion wave function. Hence all the foregoing remarks extend to the case of higher values of the spin as well, which agrees with our general formulas, as the value of $P_{bdry}(\infty)$ does not depend on the spins in the isotropic limit $t\rightarrow\infty$. In terms of the free boson, the boundary condition corresponding to the Kondo strong coupling fixed point is $\phi_l-\phi_r=\sqrt{\pi\over 2}\hbox{modulo }\sqrt{2\pi}$. 

What happens in the anisotropic case is somewhat less clear, as the boundary condition left over in the strong coupling limit now seems to depend on the impurity $sl(2)_q$ spin. 

\section{Conclusion}

It is tempting to carry out a little bit further and propose a DDV equation for the double boundary sine-Gordon model.

\begin{center}
\begin{figure}
\parbox{5cm}{
\centerline{\epsfig{figure=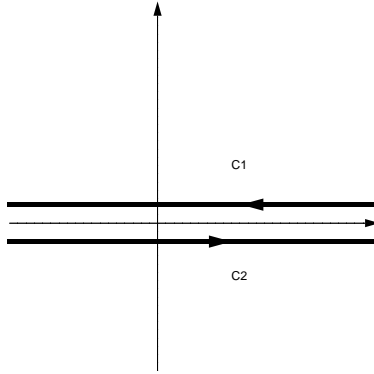,width=5cm}}}
\caption{Contours used in equation (\ref{DDVeqn}).}
\label{DDVcontour}
\end{figure}
\end{center}

To do so, we would like first to get back to the free fermion case. We introduce the two eigenvalues of the matrix $\bar{K}_lK_r$, which we call $\Lambda_\pm$. Their expressions are (setting $\mu=1$)
\begin{equation}
\Lambda_\pm={1\over (e^\theta+\Delta_l^2)(e^\theta+\Delta_r^2)}
\left[e^{2\theta}+\Delta_l^2\Delta_r^2\cos 2\chi\pm i\sqrt{\Delta_l^2\Delta_r^2 \sin^2 2\chi-e^{2\theta}(\Delta_l^2+\Delta_r^2+2\Delta_l^2\Delta_r^2\cos 2\chi)}\right]
\end{equation}
and we rewrite the ground state energy (\ref{bulkgroundstate}) as (recall $\kappa=e^\theta$) 
\begin{equation}
E=-{1\over 4\pi}\int_{-\infty}^\infty d\theta e^\theta \ln\left(1+\Lambda_+(\theta)e^{-Re^\theta}\right)+
-{1\over 4\pi}\int_{-\infty}^\infty d\theta e^\theta \ln\left(1+\Lambda_-(\theta)e^{-Re^\theta}\right)
\end{equation}
We now perform a shift of the variable of integration, $\theta\rightarrow \theta+{i\pi\over 2}-i\epsilon$ in the first integral, $\theta\rightarrow \theta-{i\pi\over 2}+i\epsilon$ in the second integral. Here, the $\epsilon's$ are necessary to avoid the zeroes of the logarithms in the integral. Care must also be exercised in moving the contours because of the branch cuts in the definitions of the square roots in $\Lambda_\pm$.  This gives rise to the new result
\begin{equation}
E={1\over 4i\pi}\left[
\int_{C_1} e^\theta\ln\left(1-{1\over\lambda_+}e^{Rie^\theta}\right)
+\int_{C_2} e^\theta\ln\left(1-\lambda_+e^{-Rie^\theta}\right)\right]
\end{equation}
where we have introduced the eigenvalues of the product $R_lR_r$ (no complex conjugation, see later), 
\begin{equation}
\lambda_\pm={1\over (e^\theta-i\Delta_l^2)(e^\theta-i\Delta_r^2)}
\left[-e^{2\theta}+\Delta_l^2\Delta_r^2\cos 2\chi\pm i\sqrt{\Delta_l^2\Delta_r^2 \sin^2 2\chi+e^{2\theta}(\Delta_l^2+\Delta_r^2+2\Delta_l^2\Delta_r^2\cos 2\chi)}\right]
\label{DDVeqn}
\end{equation}
and the contours are represented in the figure \ref{DDVcontour}. 

Similar manipulations could lead to the same result with $\lambda_+$ replaced with $\lambda_-$, and the two are in fact equal. After some simple redefinitions we thus obtain the result
\begin{equation}
E={1\over 4i\pi} \left[\int_{C_1} e^\theta \ln\left(1-e^{f(\theta)}\right)
+\int_{C_2} e^\theta \ln\left(1-e^{-f(\theta)}\right)
\right]
\end{equation}
where 
\begin{equation}
f(\theta)=Rie^\theta-\ln\lambda_\pm
\end{equation}

We conjecture that the DDV expression in the general case is very similar, with the exact same form for the energy, and as for the source equation
\begin{equation}
f(\theta)=Rie^\theta-\ln\lambda_++\int_{C_1}\Phi(\theta-\theta')\ln\left(1-e^f\right)+\int_{C_2}\Phi(\theta-\theta')\ln\left(1-e^{-f}\right)
\end{equation}
and $\Phi$ is the kernel already used in the DDV equations of the previous paragraph. 

Here, $\lambda_\pm$ are eigenvalues of $R_lR_r$, which have a structure entirely similar to the one for the case 
$\lambda=1$, 
\begin{eqnarray}
\lambda_\pm ={1\over\prod_{j=1}^\lambda \left(e^\theta-\Delta_l^{\lambda+1/
\lambda}e^{i\pi/2\lambda}e^{-i\pi j/\lambda}\right)
\left(e^\theta-\Delta_r^{\lambda+1/
\lambda}e^{i\pi/2\lambda}e^{-i\pi j/\lambda}\right)} \times \nonumber\\
\times \left[e^{2\lambda\theta}-\Delta_l^{\lambda+1}\Delta_r^{\lambda+1}\cos\eta
\pm i\left[\Delta_l^{2\lambda+2}\Delta_r^{2\lambda+2}\sin^2\eta
+e^{2\lambda\theta}\left(\Delta_l^{2\lambda+2}\Delta_r^{2\lambda+2}+2\cos\eta\Delta_l^{\lambda+1}\Delta_r^{\lambda+1}\right)\right]^{1/2}\right]
\end{eqnarray}
(and recall $\eta=(\lambda+1)\chi$). An important property is that $\lambda_+(\theta+i\pi)=\bar{\lambda}_-(\theta)$. 

It is interesting to see what our conjecture gives in the case where $\beta^2\rightarrow 8\pi$. There one finds
\begin{equation}
\lambda_\pm={-1-\Delta_l\Delta_r\cos\chi\pm
 i\sqrt{\Delta_l^2\Delta_r^2\sin^2\chi+\Delta_l^2+\Delta_r^2-2\Delta_l\Delta_r\cos\eta}\over\sqrt{(1+\Delta_l^2)(1+\Delta_r^2)}}
\end{equation}
In this limit, the dependence on the energy of the extra soliton fugacity has disappeared, and the ground state energy follows from the usual dilogarithm calculation
\begin{equation}
E=-{\pi\over 24 R}\left(1-{6\over\pi}\alpha^2\right)
\end{equation}
The eigenvalues being pure phases, one has 
\begin{equation}
\alpha=\pm \hbox{arg }\lambda_\pm =\mp \arctan {\sqrt{\Delta_l^2\Delta_r^2\sin^2\chi+\Delta_l^2+\Delta_r^2-2\Delta_l\Delta_r\cos\chi}\over
1+\Delta_l\Delta_r\cos\chi}
\end{equation}
The relation between the coupling $\Delta$ and the bare coupling is simply $\Delta=\pi\tilde{\Delta}$. 

We check this expression by comparing the resulting ``current'' in applications to Josephson juncionts to the one obtained in \cite{AffleckPRB62}. The current follows
\begin{equation}
I\propto \alpha{d\alpha\over d\chi}\propto \Delta_l\Delta_r\sin\chi{\alpha\over \sqrt{\Delta_l^2\Delta_r^2\sin^2\chi+\Delta_l^2+\Delta_r^2-2\Delta_l\Delta_r\cos\chi}}
\end{equation}
and $\alpha$ can be written
\begin{equation}
\alpha=
\arccos{1+\Delta_l\Delta_r\cos\chi\over \sqrt{(1+\Delta_l^2)(1+\Delta_r^2)}}
\end{equation}
It is easy to check that this coincides with the formula in \cite{AffleckPRB62} but for the redefinition
\begin{equation}
\Delta_{here}={\Delta_{there}\over 1-\Delta_{there}^2}
\end{equation}

Getting back to arbitrary values of $\lambda$ 
, in the limit $R\rightarrow\infty$, it is easy to check from the same kind of calculations that the correct value
\begin{equation}
E_0=-{\pi\over 24R} \left(1-6t{\chi^2\over\pi^2}\right)
\end{equation}
is recovered. 

Of course, these few tests are far from enough to completely justify our proposal, and we expect to get back to it in more details in the future.

As we were completing this paper, a preprint appeared  \cite{Leeh0301075} 
where the same problem is discussed, and the problem of analytic continuation is also claimed to be solved. We believe that the solution proposed 
in \cite{Leeh0301075} is actually incorrect, for technical 
reasons we now discuss.

 In this preprint, the authors claim to 
 find the solutions to the singularity conditions on the imaginary
axis of $\kappa$, for any value of $\lambda$.  While this is certainly true for
$\lambda = 1$ as shown in our earlier work \cite{CauxPRL88}, we have shown
that for $\lambda = 2$, the singularities should be
found on the lines with angles $\pm \pi/4$ in the complex plane of $\kappa$, instead of the imaginary axis.
Moreover, the authors of \cite{Leeh0301075} argue that only one singularity shows up for any value of 
$\lambda$, whereas we have  shown (using for example of $\kappa 
\rightarrow 0$ asymptotics and the $Y$-system arguments) that two 
singularities have to be treated for
$\lambda = 2$ to ensure analyticity in $\chi$.   
As an illustration, consider  equation (3-12) of \cite{Leeh0301075}, which
is the universal form of the TBA corrected for the first singularity.  The
result for the correction is the residue $\ln S_{a0}$, whereas we correct
with the residue of the universal scattering kernel (see our equation
(\ref{univcorrected})).  The reason for the 
difference seems to be that the authors of \cite{Leeh0301075} have not taken into
account the fact that the function $\epsilon_1$ diverges at $\kappa = 0$
when $\chi = \pi/3$, in addition to the first logarithmic kernel 
involving $\epsilon_2$ (the notations are different:  $\epsilon_0$ in their
work is what we have called $\epsilon_2$ here).  
These two divergences each modify the TBA equations,
and correcting for only one of them leads to the wrong results.  This is
transparent if the whole analytical continuation procedure is done using the
universal for the the TBA equations from the start, as we have done.

\begin{acknowledgements}
We thank P. Dorey, R. Egger and  A. Zamolodchikov for helpful conversations. 
This work was supported by Stichting FOM, the DOE and the Humboldt Foundation (HS). 
\end{acknowledgements}

\begin{appendix}

\section{The free fermion point}

At $\beta^2 = 4\pi$, the sine-Gordon action has as is well-known a
direct correspondence to the action of free fermions.
In order to correctly treat our double-boundary theory at this free fermion
point, we have to be very careful with our bosonization/refermionization
identities.  These are much simpler in the bulk, where we can simply 
toss away the finite-size terms, and for simple periodic boundary 
conditions, for which the quantizations rules of the zero modes can be
directly read out.  We will start from the very basics to
illustrate the procedure forced upon us by the two-boundary geometry
we are considering, and give every detail of the derivation from the
canonical quantization of the free boson to the final formula for the
ground-state energy.

Consider thus the interval $I: x \in [0,R]$ on the real line.  We want to
study a massless bosonic field $\phi(x) $ which is free in the bulk 
(i.e. for $x \in
~]0,R[$) but has some boundary contribution to the action 
at the points $x = 0,R$.  If we intend to develop $\phi(x)$ in a
mode expansion, we might very well be tempted to use periodicity with 
period $R$.  As we do not wish to identify $\phi(x)$ at the points 
$0$ and $R$, in order to accommodate different boundary effects on
both sides, this is not flexible enough.
Instead, we will extend the definition of $\phi(x)$ to the interval
$x \in [0,2R]$ by using 
\begin{eqnarray}
\phi(2R -x) = \phi(x), \hspace{3cm} x \in [0, R].
\label{extension}
\end{eqnarray}
We can then consistently extend the definition of $\phi(x)$ to the 
entire real line by using
\begin{eqnarray}
\phi(x+2R) = \phi(x), \hspace{1cm} x \in {\Bbb R}.
\label{continuation}
\end{eqnarray}
This is important:  the requirement that the fields on both ends of
the original interval were not identified with one another, has 
required us to define the field as periodic on an interval with 
${\it double}$ the original length.
 
Let us consider the free Hamiltonian
\begin{eqnarray}
H_0 = \frac{1}{2} \int_0^R dx [(\partial_t \phi)^2 + (\partial_x
\phi)^2].
\end{eqnarray}
Imposing the properties of $\phi$ in (\ref{extension},\ref{continuation}), 
we can write the mode expansion
\begin{eqnarray}
\phi(x,t) = \phi^0 + \Pi_0 \frac{t}{R} + \frac{i}{\sqrt{\pi}}
\sum_{n\neq 0} \frac{1}{n} a_n \cos \frac{\pi n x}{R} e^{-i \pi n t/R}
\end{eqnarray}
where the commutation relations of the modes read
\begin{eqnarray}
[\phi^0, \Pi_0] = i, \hspace{3cm} [a_n, a_m] = n \delta_{n+m, 0}
\end{eqnarray}
with all others vanishing.  In terms of these, the Hamiltonian reads
\begin{eqnarray}
H_0 = \frac{\Pi_0^2}{2R} + \frac{\pi}{2R} \sum_{n\neq 0} a_n a_{-n}.
\label{Hamiltonianmodes}
\end{eqnarray}
Alternately, we can play another game and
define a ${\it chiral}$ left-moving boson as
\begin{eqnarray}
\phi_L(t+x) = \phi_L^0 + \Pi_0 \frac{t+x}{2R} + \frac{i}{\sqrt{4\pi}} \sum_{n \neq 0} \frac{1}{n} a_n
e^{-i\pi n(t+x)/R}
\end{eqnarray}
obeying the quasi-periodicity relation
\begin{eqnarray}
\phi_L(x+2R) = \phi_L(x) + \Pi_0.
\end{eqnarray}
We can without contradiction do a nonlocal identification between our
original bosonic field and this newly invented left-mover.  Namely, we
may at leisure impose the operator identity
\begin{eqnarray}
\phi(x,t) = \phi_L(t+x) + \phi_L(t-x) + \phi^0 -2\phi_L^0
\end{eqnarray}
where the periodicity requirements of the original boson are automatically
fulfilled given the quasi-periodicity of the chiral one.
Note that under these circumstances, 
the Hamiltonian (\ref{Hamiltonianmodes}) becomes
\begin{eqnarray}
H_0 = \int_0^{2R}dx (\partial_x \phi_L)^2.
\end{eqnarray}
Moreover, it is a trivial exercise to show that 
taking $\phi_L^0 = \phi^0/2$ leads to consistent canonical commutation relations for the
chiral field:
\begin{eqnarray}
[\phi_L(x), \partial_{x'}\phi_L(x')] = \frac{i}{2} \sum_{n \in \Bbb{Z}} \delta (x-x'+2Rn).
\end{eqnarray}

Let us now move on to the model we are interested in.  
At the free fermion point, the boundary contributions to the Hamiltonian 
read
\begin{eqnarray}
H_B = \tilde{\Delta}_l \cos \sqrt{\pi}\phi(0,t) + \tilde{\Delta}_r
\cos (\sqrt{\pi}\phi(R,t)  -\chi).
\label{boundaryH}
\end{eqnarray}
To refermionize this at $\beta^2 = 4\pi$, we start by noting that the 
original boson at $x = 0,R$ is simply expressed in terms of the chiral
boson:
\begin{eqnarray}
\phi(0,t) = 2 \phi_L (t), \hspace{1cm} 
\phi(R,t) = 2 \phi_L (R + t)  -\Pi_0
\end{eqnarray}
Progress is now simply a matter of using the definition of the chiral fermion
from our favourite bosonization/refermionization rules,
\begin{eqnarray}
\Psi = \frac{1}{\sqrt{4\pi a}} e^{-i\sqrt{4\pi} \phi_L},
\end{eqnarray}
to obtain the equivalent fermionic boundary Hamiltonian
\begin{eqnarray}
H_B = {\Delta_l\over \sqrt{2}} [\Psi(0)+\Psi^{\dagger}(0)] + i {\Delta_r\over \sqrt{2}} [f \Psi(R) e^{i\chi}+
f^{\dagger} \Psi^{\dagger}(R) e^{-i\chi}]
\end{eqnarray}
where we have defined $f = e^{-i\sqrt{\pi} \Pi_0}$ and rescaled our
boundary parameters as $\Delta = \sqrt{\pi} ~\tilde{\Delta}$, setting the cut-off $a$ equal to one.
Note that we now 
face a linear fermionic term at $x=0$, and that the behaviour of the
operator $\Pi_0$, and thus that of $f$, is still unspecified.
To resolve this is simply a matter of noting that 
the operator $\Pi_0$ becomes quantized through the periodicity requirement that 
we imposed on the original theory.  Writing the boundary contributions in
terms of the left-mover, we get the $\Pi_0$ 
quantization relation and (being very careful with noncommuting 
operators) its fermionic implication
\begin{eqnarray}
\Pi_0 = \sqrt{\pi} n, \hspace{2cm} \Psi(x +2R) = -\Psi(x)
\end{eqnarray}
with $n \in {\Bbb Z}$.  These then tell us that 
the operator $f$ in fact obeys the Majorana relations
\begin{eqnarray}
f = f^{\dagger}, \hspace{1cm} \{f, f\} = 2, \hspace{1cm} \{f, \Psi \} = 0.
\end{eqnarray}
For aesthetic reasons, we would like our Hamiltonian to be
quadratic in fermions, and somehow more symmetric with
respect to the left and right components.  In order to do this,
we perform the simple transformation
\begin{eqnarray}
\Psi \rightarrow -a \Psi, \hspace{1cm} f \rightarrow i ~a ~b
\end{eqnarray}
where $a, b$ are two new Majorana fermions, living respectively on the left and right
boundaries. One can check that all anticommutators are preserved under this
transformation.

We thus finally obtain a better-shaped real-time action,
\begin{eqnarray}
S &=& \int dt \int_0^{2R} dx \left[ \frac{1}{2} \Psi^{\dagger} i (\partial_t -\partial_x) \Psi + 
\Psi i (\partial_t -\partial_x) \Psi^{\dagger} \right] + 
\nonumber \\
&&+\int dt \biggl[\frac{i}{2}(a \partial_t a + b\partial_t b) 
 +{\Delta_l\over \sqrt{2}} a  [\Psi(0)-\Psi^{\dagger}(0)] +  
{\Delta_r\over \sqrt{2}} b [\Psi (R)e^{i \chi} - 
\Psi^{\dagger} (R)e^{-i\chi}]  \biggr].
\end{eqnarray}

Our theory has thus boiled down to something rather obviously tractable, 
namely one that is quadratic in fermions, and which we can consequently solve
exactly.  
Varying the action and eliminating the boundary fermions, we obtain two sets of
boundary conditions at $x=0$ and $R$:
\begin{eqnarray}
&&\Psi(0^+,t) + \Psi^{\dagger}(0^+,t) = \Psi(0^-,t) +  \Psi^{\dagger}(0^-,t)  \\
\nonumber
\\ 
&&\partial_t \Psi(0^+,t) -\partial_t \Psi^{\dagger}(0^+,t) - \partial_t \Psi(0^-,t) +
\partial_t \Psi^{\dagger}(0^-,t)= \nonumber \\
&&\hspace{1cm} ={\Delta_l^2\over 2} \left[ \Psi(0^+,t) - \Psi^{\dagger}(0^+,t) 
+ \Psi(0^-,t) - \Psi^{\dagger}(0^-,t)
\right]
\\ \nonumber \\ 
&&\Psi(R^+,t) + e^{-2i\chi}\Psi^{\dagger}(R^+,t) = \Psi(R^-,t) + e^{-2i\chi} 
\Psi^{\dagger}(R^-,t) \\ \nonumber \\ 
&&\partial_t \Psi(R^+,t) - e^{-2i\chi} \partial_t \Psi^{\dagger}(R^+,t) -\partial_t \Psi(R^-,t) 
+ e^{-2i\chi} \partial_t \Psi^{\dagger}(R^-,t)= \nonumber \\
&&\hspace{1cm}={\Delta_r^2\over 2} \left[ \Psi(R^+,t) - e^{-2i\chi} \Psi^{\dagger}(R^+,t) 
+ \Psi(R^-,t) - e^{-2i\chi}
\Psi^{\dagger}(R^-,t) \right]
\label{boundaryconditions}
\end{eqnarray}

The fermion is a free left-mover in the bulk, with possible discontinuities at $x = 0,R$.  We
choose the general mode expansions
\begin{eqnarray}
\Psi(x,t) &=& \sum_k c_ke^{-ik(x+t)} \hspace{3cm} 0<x<R \nonumber \\
\Psi(x,t) &=& \sum_k d_ke^{-ik(x+t)} \hspace{3cm} R<x<2R
\end{eqnarray}
and substitute them back into (\ref{boundaryconditions}), keeping in mind the 
antiperiodicity 
of the fermions under $x \rightarrow x+2R$.
This scheme is consistent
provided the momenta $k$ are quantized according to
\begin{eqnarray}
1 + W &=& 0, \nonumber \\
W &=& 2\frac{\left[4(ik)^2 + \Delta_l^2 \Delta_r^2 \cos 2\chi
\right]}{[2ik + \Delta_l^2][2ik+\Delta_r^2]}
e^{-2ikR} +\frac{[2ik-\Delta_l^2][2ik-\Delta_r^2]}{[2ik +
\Delta_l^2][2ik+\Delta_r^2]} e^{-4ikR}
\label{quantization}
\end{eqnarray}

\begin{center}
\begin{figure}
\parbox{8cm}{
\epsfig{figure=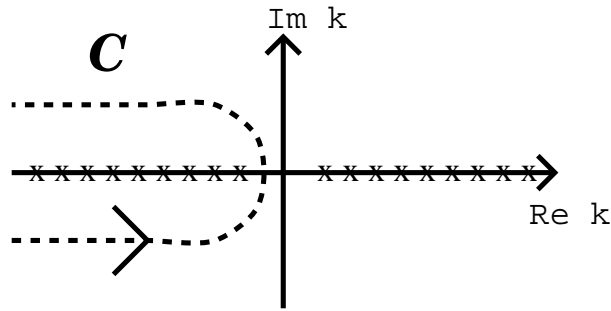,width=8cm}}
\vspace{0.3cm}
\caption{Contour of integration for performing the summation over
occupied energy states when calculating the ground-state energy.  The
crosses represent allowed values of the quasimomentum, which are
symmetrically distributed around $k=0$.}
\label{fig:Contour}
\end{figure}
\end{center}

The ground-state energy is then given by 
\begin{eqnarray}
E_0 = \frac{1}{2} \sum_{k < 0} k = \frac{1}{4\pi i} \int_{\cal C} d k k \frac{\frac{dW}{dk}}{1+W}
= -\frac{1}{4\pi i} \int_{\cal C} dk \ln ~[1 + W].
\end{eqnarray}
where the anti-clockwise contour ${\cal C}$ surrounds all the roots of our quantization 
condition (\ref{quantization}) lying on the negative half-line $k <
0$.  This is illustrated in figure \ref{fig:Contour}.

The contour ${\cal C}$ can be split in two different parts, $C_{1,2}$,
corresponding respectively to the upper and lower branches.  The first
integral (over $C_1$) then converges after performing a Wick rotation
$\kappa = ik$, whereas the second integral (over $C_2$) converges
after extracting a multiplicative factor
$\frac{[2ik-\Delta_l^2][2ik-\Delta_r^2]}{[2ik + 
\Delta_l^2][2ik+\Delta_r^2]} e^{-4ikR}$ from within the argument of the
logarithm while performing the Wick rotation $\kappa = -ik$.  Setting
the nonuniversal term $\int_0^{\infty} dk k$ to zero, we then obtain
the final formula
\begin{eqnarray}
E_0 = -\frac{1}{4\pi} \int_0^{\infty} d \kappa \ln \left[ 1+2
\frac{\kappa^2 + \Delta_l^2 
\Delta_r^2 \cos 2\chi}{(\kappa +  \Delta_l^2)(\kappa +  \Delta_r^2)}
e^{-\kappa R} 
+ 
\frac{(\kappa -  \Delta_l^2)(\kappa -  \Delta_r^2)}{(\kappa +  \Delta_l^2)(\kappa
+  \Delta_r^2)} e^{-2\kappa R}
\right]  +\epsilon_l + \epsilon_r
\label{groundstate1}
\end{eqnarray}
where $\epsilon_{l,r}$ are the boundary intensive energies
\begin{eqnarray}
\epsilon_{l,r} = -\frac{1}{4\pi} \int_0^{\infty} d k \frac{4 k
\Delta_{l,r}^2}{4k^2 +\Delta_{l,r}^4}. 
\end{eqnarray}
We have thus obtained the ground-state energy of the $\beta^2 = 4\pi$
sine-Gordon model in a finite width $R$, in the presence of two
integrable boundary interactions.  Although the integral is not
expressible in terms of elementary functions, it gives us the
dependence of $E_0$ in terms of the boundary parameters $\Delta_l,
\Delta_r$ and phase $\chi$.

\section{Bulk sine-Gordon scattering matrices}

In the bulk, the scattering of the soliton and antisoliton is
described by the general processes
\begin{eqnarray}
A_1^{\dagger}(\theta_1) A_1^{\dagger} (\theta_2) = a(\theta_1 - \theta_2)
 A_1^{\dagger}(\theta_2) A_1^{\dagger} (\theta_1),  \nonumber \\
A_2^{\dagger}(\theta_1) A_2^{\dagger} (\theta_2) = 
a(\theta_1 - \theta_2)
A_2^{\dagger}(\theta_2) A_2^{\dagger} (\theta_1),  \nonumber \\
A_1^{\dagger}(\theta_1) A_2^{\dagger} (\theta_2) = b(\theta_1 - \theta_2)
A_2^{\dagger}(\theta_2) A_1^{\dagger} (\theta_1) + c(\theta_1 -
 \theta_2) A_1^{\dagger}(\theta_2) A_2^{\dagger} (\theta_1).
\end{eqnarray}
The amplitudes for these scattering processes are well-known 
\cite{ZamolodchikovAP120}.  When $\beta^2 = \frac{8\pi}{\lambda+1}, 
~~\lambda  \in
{\Bbb N}$, the theory contains the following fundamental particles:
the soliton, the antisoliton, and $\lambda -1$ breathers of mass
$m_n = 2m \sin \frac{\pi n}{2 \lambda}$.  The bulk scattering simplifies
since  the backscattering amplitude $c(\theta)$ vanishes;  the
other two amplitudes read
\begin{eqnarray}
a(\theta) &=& i \sinh [\lambda(\theta - i \pi)] \rho(\theta), \nonumber \\
b(\theta) &=& -i \sinh [\lambda \theta] \rho (\theta), \nonumber \\
\rho(\theta) &=& -\frac{1}{\pi}  \Gamma(\lambda) \Gamma(1 + i
\frac{\lambda \theta}{\pi}) \Gamma(1-\lambda - i \frac{\lambda
\theta}{\pi}) 
\prod_{l=1}^{\infty} \frac{F_l(\theta) F_l(i\pi - \theta)}{F_l(0)
F_l(i \pi)}, \nonumber \\
F_l (\theta) &=& \frac{\Gamma(2l\lambda+ i \frac{\lambda \theta}{\pi})
\Gamma(1 + 2l\lambda + i \frac{\lambda \theta}{\pi})}
{\Gamma((2l+1) \lambda+ i \frac{\lambda \theta}{\pi})
\Gamma(1 + (2l-1)\lambda + i \frac{\lambda \theta}{\pi})}
\end{eqnarray}
A useful integral representation for this is \cite{FendleyNPB430}
\begin{eqnarray}
\rho(\theta) = \frac{1}{\sin [\lambda(\pi + i \theta)]} \exp 
\left( i \int_{-\infty}^{\infty} \frac{dy}{2y} \sin{\frac{2\lambda
\theta y}{\pi}} \frac{\sinh[(1-\lambda)y]}{\sinh{y}~{\cosh{\lambda y}}}
\right).
\end{eqnarray}
At the reflectionless points $\beta^2 = \frac{8\pi}{\lambda +1},
\lambda \in {\Bbb N}$, the bulk scattering amplitudes simplify
considerably.  Namely, the backscattering amplitude $c(\theta)$
vanishes identically, while the others simplify according to
\begin{eqnarray}
a(\theta) = - \prod_{j=1}^{\lambda} \frac{\cos \left( \frac{\pi j}{2
\lambda} + i\frac{\theta}{2} \right)}{\cos \left( \frac{\pi j}{2
\lambda} - i\frac{\theta}{2}\right)}, \hspace{2cm} b(\theta) = \frac
{ \sin (-i \lambda \theta)}{\sin( \lambda(\pi + i\theta))} a(\theta) = (-1)^{\lambda + 1}
a(\theta).
\end{eqnarray}

These have to be supplemented with the scattering amplitudes for 
processes involving breathers, which are defined as
\begin{eqnarray}
A^{\dagger} (\theta_1) B^{\dagger}_n (\theta_2) &=& S^{(n)}(\theta_1 -
\theta_2) B^{\dagger}_n (\theta_2) A^{\dagger}(\theta_1), \nonumber \\
B^{\dagger}_n (\theta_1) B^{\dagger}_m (\theta_2) &=& S^{(n,m)}(\theta_1 -
\theta_2) B^{\dagger}_m (\theta_2) B^{\dagger}_n(\theta_1), \nonumber \\
S^{(n)}(\theta) &=& \frac{\sinh \theta + i \cos \frac{\pi n}{2\lambda}}
{\sinh \theta - i \cos \frac{\pi n}{2\lambda}} \prod_{l=1}^{n-1}
\frac{\sin^2 (\frac{(n-2l)\pi}{4 \lambda} - \frac{\pi}{4} + i
\frac{\theta}{2})}{\sin^2 (\frac{(n-2l)\pi}{4\lambda} - \frac{\pi}{4} - i
\frac{\theta}{2})}, \nonumber \\
S^{(n,m)}(\theta) &=& \frac{\sinh \theta + i
\sin(\frac{(n+m)\pi}{2\lambda})}{\sinh \theta - i
\sin(\frac{(n+m)\pi}{2\lambda})} 
\frac{\sinh \theta + i
\sin(\frac{(n-m)\pi}{2\lambda})}{\sinh \theta - i
\sin(\frac{(n-m)\pi}{2\lambda})}
\times \nonumber \\
&&\times \prod_{l=1}^{m-1} \frac{\sin^2 (\frac{(m-n-2l)\pi}{2\lambda} +
i\frac{\theta}{2}) \cos^2 (\frac{(m+n-2l)\pi}{2\lambda} + i
\frac{\theta}{2})}{\sin^2 (\frac{(m-n-2l)\pi}{2\lambda} -
i\frac{\theta}{2}) \cos^2 (\frac{(m+n-2l)\pi}{2\lambda} - i
\frac{\theta}{2})}, \hspace{1cm} n \geq m.
\end{eqnarray}
The integral representations of these amplitudes read
\begin{eqnarray}
S^{(n)} (\theta) &=& \exp \left( -i \int_{-\infty}^{\infty} \frac{dy}{2y}
\sin(\frac{2\lambda \theta y}{\pi}) \frac{\sinh ny \cosh y}{\cosh
\lambda y ~\sinh y} \right), \nonumber \\
S^{(n,m)} (\theta) &=& \exp \left( i \int_{-\infty}^{\infty} 
\frac{dy}{y} \sin(\frac{2\lambda \theta y}{\pi}) \left[
\delta_{n,m} - 2 \frac{\cosh y \cosh[(\lambda -n)y] \sinh my}{\cosh
\lambda y ~\sinh y} \right] \right), \nonumber \\
&&n \geq m = 1,...,\lambda -1.
\end{eqnarray}

\section{Boundary sine-Gordon scattering matrices}

\subsection{Solution to the boundary bootstrap}
The amplitudes for the scattering of the soliton and antisoliton on
the boundaries were computed in \cite{GhoshalIJMPA9}.  
The bootstrap solution yields these amplitudes in terms of two
parameters $\eta, \vartheta$ whose relation to the parameters in the original
Lagrangian was not specified in \cite{GhoshalIJMPA9}.  For now, we
state the results in terms of $\eta, \vartheta$.  
For general $\lambda$, the amplitudes
for soliton and antisoliton boundary scattering read
\begin{eqnarray}P_{\pm}(\theta) &=& \left(\cos (-i \lambda \theta) \cos
\eta 
\cosh \vartheta \mp \sin(-i \lambda \theta) \sin \eta \sinh \vartheta
\right) R(\theta), \nonumber \\
Q_{\pm}(\theta) &=& -\frac{1}{2} \sin(-2i\lambda \theta) R(\theta),
\end{eqnarray}
where
\begin{eqnarray}
R(\theta) = R_0 (\theta) R_1 (\theta).
\end{eqnarray}
The first of these functions is given by
\begin{eqnarray}
R_0(\theta) = \frac{F_0(\theta)}{F_0(-\theta)}, 
\end{eqnarray}
where
\begin{eqnarray}
F_0 (\theta) = \frac{\Gamma(1 +i \frac{2\lambda \theta}{\pi})}
{\Gamma(\lambda +i \frac{2\lambda \theta}{\pi})}  \prod_{k=1}^{\infty}
\frac{\Gamma(4k\lambda +i \frac{2\lambda \theta}{\pi})}
{\Gamma((4k+1)\lambda  +i \frac{2\lambda \theta}{\pi})} 
\frac{\Gamma(1+ 4k\lambda +i \frac{2\lambda\theta}{\pi})
\Gamma((4k+1)\lambda) \Gamma(1+ (4k-1)\lambda)}{\Gamma(1+ (4k-1)\lambda +
i \frac{2\lambda\theta}{\pi})
\Gamma(1+4k\lambda) \Gamma(4k\lambda)}.
\end{eqnarray}
Its integral representation reads
\begin{eqnarray}
R_0 (\theta) = \exp \left( -i \int_{-\infty}^{\infty} \frac{dy}{y}
\sin (\frac{2\lambda \theta y}{\pi} )\frac{\sinh ((\lambda-1)y/2)
\sinh (3\lambda y/2)}{\sinh (y/2) \sinh (2\lambda y)} \right).
\end{eqnarray}
The second function reads
\begin{eqnarray}
R_1 (\theta) = \frac{\sigma(\eta, \theta)}{\cos \eta} \frac{\sigma(i
\vartheta, \theta)}{\cosh \vartheta}  
\end{eqnarray}
where
\begin{eqnarray}
\sigma(x, \theta) &=& \frac{\Pi(x, -\theta+i\frac{\pi}{2}) \Pi(-x,
-\theta+i\frac{\pi}{2}) \Pi(x, \theta-i\frac{\pi}{2}) \Pi(-x,
\theta-i\frac{\pi}{2})}
{\Pi^2(x, i\frac{\pi}{2}) \Pi^2(-x, -i\frac{\pi}{2})}, \nonumber \\
\Pi(x, \theta) &=& \prod_{l=0}^{\infty} \frac{\Gamma(\frac{1}{2}+(2l+
\frac{1}{2})\lambda +
\frac{x}{\pi} + i \frac{\lambda \theta}{\pi}) \Gamma(\frac{1}{2} + (2l+
\frac{3}{2})\lambda
+ \frac{x}{\pi})}{\Gamma(\frac{1}{2} +(2l+\frac{3}{2})\lambda +
\frac{x}{\pi} + i \frac{\lambda \theta}{\pi}) \Gamma(\frac{1}{2} + (2l+
\frac{1}{2})\lambda
+ \frac{x}{\pi})}.
\end{eqnarray}
This obeys the property
\begin{eqnarray}
\sigma (x, \theta) \sigma (x, -\theta) = \frac{\cos^2 x}{\cos (x+ i
\lambda \theta) \cos (x-i \lambda \theta)}
\end{eqnarray}
(correcting the equation in \cite{GhoshalIJMPA9}).  The integral
representation for these is
\begin{eqnarray}
\sigma (x, \theta) = \exp \left( i \int_{-\infty}^{\infty}
\frac{dy}{y} \sinh (\lambda y (1+\frac{i\theta}{\pi})) \sin
(\frac{\lambda \theta y}{\pi}) \frac{\cosh (\frac{2 x y}{\pi})}{\sinh
y \cosh (\lambda y)} \right).
\end{eqnarray}
Again, at the reflectionless points $\beta^2 = \frac{8\pi}{\lambda +
1}, \lambda {\Bbb N}$, these simplify considerably:
\begin{eqnarray}
R_0 (\theta) &=& \prod_{j=1}^{\lambda -1} \frac{\cos \left
( \frac{\pi}{2} + \frac{\pi j}{4\lambda} + i \frac{\theta}{2}
\right)}{\cos \left
( \frac{\pi}{2} + \frac{\pi j}{4\lambda} - i \frac{\theta}{2}
\right)}, \nonumber \\
\sigma (x, \theta) &=& \frac{\cos x}{\cos (x-i\lambda \theta)}
\prod_{j=1}^{\lambda} \frac{\cos \left(\frac{\pi}{4\lambda} -
\frac{x}{2\lambda} - \frac{\pi j}{2\lambda} + i \frac{\theta}{2}
\right)}{\cos \left(\frac{\pi}{4\lambda} -
\frac{x}{2\lambda} - \frac{\pi j}{2\lambda} - i \frac{\theta}{2}
\right)}.
\end{eqnarray}
In the massless limit, we get 
\begin{eqnarray}
R_0 (i\frac{\pi}{2} - \theta) &=& e^{-i\frac{3\pi}{4}(\lambda -1)}
\nonumber \\
\frac{\sigma (\eta, i\frac{\pi}{2}-\theta)}{\cos \eta} &=& 2 e^{-\lambda \theta
- i\pi \lambda/2} \nonumber \\
\frac{\sigma (i \vartheta, i \frac{\pi}{2}-\theta)}{\cosh \vartheta} &=&
\frac{2^{-\lambda + 1} e^{-\lambda \theta/2
-\vartheta/2}}{\prod_{j=1}^{\lambda} \cosh (\frac{\theta}{2} -
\frac{\vartheta}{2\lambda} - i\frac{\pi}{4\lambda} +i \frac{\pi
j}{2\lambda} -i\frac{\pi}{4})} \nonumber \\
R(i\frac{\pi}{2} - \theta) &=& \frac{4 e^{-i\frac{3\pi}{4}(\lambda-1) -\lambda
\theta -i \frac{\pi}{2} \lambda}}{\prod_{j=1}^{\lambda} e^{\theta} +
e^{\frac{\vartheta}{\lambda} + i \frac{\pi}{2\lambda} -i \frac{\pi
j}{\lambda} + i \frac{\pi}{2}}}.
\end{eqnarray}

The breathers also scatter off the boundary, according to
\begin{eqnarray}
B^{\dagger}_n (\theta) B = R^{(n)}_B (\theta) B^{\dagger}_n (-\theta)
B.
\end{eqnarray}
Since the breathers are bound states of (multiple) solitons and
antisolitons, their boundary scattering amplitudes can be worked out
systematically.  This procedure was carried out in
\cite{GhoshalIJMPA9bis}, with the result
\begin{eqnarray}
R^{(n)}_B(\theta) = R^{(n)}_0 (\theta) R^{(n)}_1 (\theta), 
\end{eqnarray}
where
\begin{eqnarray}
R^{(n)}_0 (\theta) = - \frac{\cos(\frac{n \pi}{4\lambda} - i
\frac{\theta}{2}) \cos(\frac{n\pi}{4\lambda} + \frac{\pi}{4} + i
\frac{\theta}{2}) \sin(\frac{\pi}{4} - i \frac{\theta}{2})}
{\cos(\frac{n \pi}{4\lambda} + i
\frac{\theta}{2}) \cos(\frac{n\pi}{4\lambda} + \frac{\pi}{4} - i
\frac{\theta}{2}) \sin(\frac{\pi}{4} + i \frac{\theta}{2})} 
\prod_{l=1}^{n-1} \frac{\sin(\frac{l \pi}{2\lambda} - i \theta)
\cos^2 (\frac{l \pi}{4\lambda} + \frac{\pi}{4} + i \frac{\theta}{2})}
{\sin(\frac{l \pi}{2\lambda} + i \theta)
\cos^2 (\frac{l \pi}{4\lambda} + \frac{\pi}{4} - i \frac{\theta}{2})}
\end{eqnarray}
and the function $R^{(n)}_1 (\theta)$ depends on the parity of $n$,
i.e.
\begin{eqnarray}
R^{(2n)}_1(\theta) &=& S^{(2n)}(\eta, \theta) S^{(2n)}(i \vartheta,
\theta), \hspace{1cm} n = 1,2, ... < \lambda/2, \nonumber \\
S^{(2n)}(x, \theta) &=& \prod_{l=1}^n \frac{\cos(\frac{x}{\lambda} -
\frac{(l-\frac{1}{2})\pi}{\lambda} ) + i \sinh \theta
}{\cos(\frac{x}{\lambda} -
\frac{(l-\frac{1}{2})\pi}{\lambda} ) - i \sinh \theta }
\frac{\cos(\frac{x}{\lambda} +
\frac{(l-\frac{1}{2})\pi}{\lambda} ) + i \sinh \theta
}{\cos(\frac{x}{\lambda} +
\frac{(l-\frac{1}{2})\pi}{\lambda} ) - i \sinh \theta } 
\end{eqnarray}
or
\begin{eqnarray} 
R^{(2n-1)}_1 (\theta) &=& S^{(2n-1)}(\eta,\theta) S^{(2n-1)}(i
\vartheta, \theta), \hspace{1cm} n = 1, 2, ... < (\lambda+1)/2, \nonumber \\
S^{(2n-1)}(x, \theta) &=& \frac{\cos(\frac{x}{\lambda}) + i \sinh
\theta}{\cos(\frac{x}{\lambda}) - i \sinh \theta} 
\prod_{l=1}^{n-1} \frac{\cos(\frac{x}{\lambda} - \frac{\pi l}{\lambda})
+ i \sinh \theta}{\cos(\frac{x}{\lambda} - \frac{\pi l}{\lambda})
- i \sinh \theta} \frac{\cos(\frac{x}{\lambda} + \frac{\pi l}{\lambda})
+ i \sinh \theta}{\cos(\frac{x}{\lambda} + \frac{\pi l}{\lambda})
- i \sinh \theta}.
\end{eqnarray}
The unified integral representation for this is worked out to be
\begin{eqnarray}
S^{(n)}(x, \theta) = \exp \left( i\pi - i \int_{-\infty}^{\infty}
\frac{d y}{y} \sin (\frac{2\lambda \theta y}{\pi}) \frac
{ \cosh(\frac{2xy}{\pi}) \sinh ny}{\cosh \lambda y \sinh y} \right).
\end{eqnarray}

\subsection{Parameter correspondence}

A crucial ingredient missing in the formulas above is the exact
relationship between the boundary parameters $\xi, k$ 
coming out of the bootstrap, and the physical variables $\Delta_{l,r},
\chi$ appearing in the boundary Hamiltonian we started from.  
This correspondence was worked out by A. Zamolodchikov
\cite{Zamolodchikovprivate} (see also the semiclassical treatment
in \cite{ChenaghlouIJMPA15}). 
For a problem where the bulk sine-Gordon action has a term 
$2\mu\cos\beta\phi$ and a boundary term $2\mu_B\cos\beta\phi/2$, one has 
\begin{eqnarray}
\cosh (\frac{\beta^2}{8\pi} (\vartheta \pm i \eta)) =
\frac{\mu_B}{\sqrt{\mu}} \sqrt{\sin \frac{\beta^2}{8}} e^{\pm i \chi}.
\end{eqnarray}
In the massless limit, we will in general thus get
\begin{eqnarray}
\vartheta \approx (\lambda + 1) \ln \left[2\frac{\mu_B}{\sqrt{\mu}}
\sqrt{\sin \frac{\pi}{\lambda+1}} \right], \hspace{2cm} \eta =
(\lambda + 1) \chi.
\end{eqnarray}
Now from the paper \cite{FSW} we have that, in the massles limit, ${\theta-\vartheta\over\lambda}=\theta'-\theta_B$ so $T_B\equiv {1\over 2}e^{\theta_B}=
me^{\vartheta/\lambda}$ (in the last formula, the factor ${1\over 2}$ comes from your normalizations in the massless limit, different from the ones in \cite{FSW}). Here $m$ is the soliton mass, related with the bare sine-Gordon parameter $\mu$ by \cite{Zamolodchikovmass}
\begin{equation}
\mu={\Gamma(1/\lambda+1)\over\pi\Gamma(\lambda/\lambda+1)}\left[
m\sqrt{\pi}\Gamma(1/2+1/2\lambda)\over 2\Gamma(1/2\lambda)\right]^{2\lambda/\lambda+1}
\end{equation}
Putting everything together just reproduces the relation between $T_B$ and $\mu_B$ mentioned in the text.

\end{appendix}


\begin{thebibliography}{99}

\bibitem{CardyBOOK} J. Cardy, ``Conformal invariance and surface
critical behavior'', in {\it Conformal Invariance and Applications to
Statistical Mechanics}, eds. C. Itzykson, H. Saleur and J. B. Zuber
(World Scientific, 1988).
\bibitem{LeClairNPB453} A. LeClair, G. Mussardo, H. Saleur and S. Skorik, 
Nucl. Phys. B 453 (1995), 581.
\bibitem{Itoyama} H. Itoyama, T. Oota, ``Normalization of Off-shell boundary state, g-function and Zeta function regularization'', hep-th/0206123.
\bibitem{DoreyNPB525} P. Dorey {\it et al.}, Nucl. Phys. B 525, 641
(1998). 
\bibitem{Feverati} G. Feverati, P. A. Pearce and F. Ravanini, Phys. Lett. B534 (2002) 216.
\bibitem{Takacs} Z. Bajnok, L. Palla and G. Takacs, Nucl. Phys. B622 (2002) 565.
\bibitem{Giuseppe} P. Mosconi, G. Mussardo and V. Riva, Nucl. Phys. B621 (2002) 571. 
\bibitem{Affleck} I. Affleck, Nuc. Phys. B336 (1990) 517.
\bibitem{CauxPRL88}  J.-S. Caux, H. Saleur and F. Siano, Phys. Rev. Lett. 88 (2002) 106402.
\bibitem{Bardakci} K. Bardakci and A. Konechny, ``Tachyon instability and Kondo type models'', hep-th/0009214.
\bibitem{Harvey} J. A. Harvey, D. Kutasov and E. J. Martinec, ``On the relevance of tachyons'', hep-th/0003101.
\bibitem{GhoshalIJMPA9} S. Ghoshal and A. B. Zamolodchikov, Int. Jour. 
Mod. Phys. A 9 (1994), 3841.
\bibitem{Baseilhac} P. Baseilhac and K. Koizumi, ``Sine-Gordon quantum field theory on the half-line with quantum boundary degrees of freedom'', hep-th/0208005.
\bibitem{DestriDeVega} C. Destri and H.  de Vega, Phys. Rev. Lett. 69 (1992) 2313.  
\bibitem{Klumper} P. A. Pearce and A. Kl\"umper, Phys. Rev. Lett. 68 (1991) 974; A. Kl\"umper and P. A. Pearce, J. Stat. Phys. 64 (1991) 13.
\bibitem{Zamolodchikovmass} Al. Zamolodchikov, J. Mod. Phys. A10 (1995) 1125.
\bibitem{Chatterjee} R. Chatterjee, Nucl.Phys. B468 (1996) 439.
\bibitem{FSW} P. Fendley, H. Saleur and N. Warner, Nucl. Phys. B430 (1994) 577.
 \bibitem{YangPRB11} H. Au. Yand and M. Fisher, Phys. Rev. B11, 3469
(1975). 
\bibitem{TakhtadjianTMP21} L. Takhtadjian and L. Faddeev,
Theor. Math. Phys. 21, 160 (1974);  V. Korepin and L. Faddeev,
Theor. Math. Phys. 25, 147 (1975).
\bibitem{ZamolodchikovAP120} A. B. Zamolodchikov and
Al. B. Zamolodchikov, Ann. Phys. 120, 253 (1979).
\bibitem{AmeduriPLB354} M. Ameduri, R. Konik and A. LeClair, Phys. Lett. B 
354 (1995), 376.
\bibitem{FendleyNPB430} P. Fendley, H. Saleur and N. P. Warner,
Nucl. Phys. B 430, 577 (1994).
\bibitem{GhoshalIJMPA9bis} S. Ghoshal, Int. Jour. Mod. Phys. A 9, 4801
(1994).
\bibitem{AffleckPRB62} I. Affleck, J.-S. Caux and A. M. Zagoskin, Phys. 
Rev. B 62 (2000), 1433.
\bibitem{AffLu} I. Affleck and A. W. W. Ludwig, Nucl. Phys. B352 (1991) 849.
\bibitem{AOS} I. Affleck, M. Oshikawa and H. Saleur, Nucl.Phys. B594 (2001) 535-606.
\bibitem{Zamolodchikovprivate} Al. B. Zamolodchikov, private
communication.
\bibitem{ZamolodchikovPLB253} Al. B. Zamolodchikov, Physics Letters B
253, 391 (1991).
\bibitem{ChenaghlouIJMPA15} A. Chenaghlou and E. Corrigan, Int. Jour.
Mod. Phys. A 15, 4417 (2000).
\bibitem{Leeh0301075} T. Lee and C. Rim, ``Thermodynamic Bethe ansatz for boundary sine-Gordon model'', hep-th/0301075.

\end{thebibliography}
\end{document}